\documentclass[10pt, usenatbib]{aa}
\usepackage{times}
\usepackage{graphicx}
\usepackage[varg]{txfonts}
\usepackage{mathtools}
\usepackage{subscript}
\usepackage{subcaption}
\usepackage{courier}
\bibpunct{(}{)}{,}{a}{}{,} 

\begin{document}
 
 \title{Spectral disentangling with Spectangular\thanks{Based in part on data obtained with
the STELLA robotic telescope in Tenerife, an AIP
facility jointly operated by AIP and IAC.}}
 
 \author{Daniel P. Sablowski\thanks{Corresponding author.
 \email{dsablowski@aip.de}} \& Michael Weber}
 
 \institute{Leibniz-Institute for Astrophysics Potsdam (AIP), An der Sternwarte 16, D-14482 Potsdam}
              
 \titlerunning{Spectral Disentangling with Spectangular}
 
 \authorrunning{Daniel P. Sablowski}

 \date{Received 2016; accepted 2016}
 
 
 \abstract{The paper introduces the software \texttt{Spectangular} for spectral disentangling via singular value decomposition with global optimisation of the orbital parameters of the stellar system or radial velocities of the individual observations. We will describe the procedure and the different options implemented in our program.
 Furthermore, we will demonstrate the performance and the applicability using tests on artificial data. Additionally, we use high-resolution spectra of Capella to demonstrate the performance of our code on real-world data.
 The novelty of this package is the implemented global optimisation algorithm and the graphical user interface (GUI) for ease of use. We have implemented the code to tackle SB1 and SB2 systems with the option of also dealing with telluric (static) lines.}
              
\maketitle

\titlerunning{Spectral Disentangling with Spectangular}

\sloppy

\section{Introduction}
\label{sec:intro}

Since stellar evolution depends on the mass of the object, measurement of radial velocities of binary stars is, besides the measurement of the inclination, an important step for stellar evolution studies. It has been shown, e.g.\  by \citet{2016A&A...586A..35G} and \citet{2004ASPC..318...95H}, that disentangling the spectra helps to improve determining the orbit of binaries. Furthermore, it allows detailed chemical analysis of the component spectra \citep{2005A&A...439..309P}.

Firstly, we note that decomposition methods like direct and iterative subtraction \citep[e.g.,][]{2006A&A...448..283G} are not understood as disentangling techniques. These subtraction methods rely on a fully analytical approach which is not fulfilled by real (noisy) measurements.
Their procedure is to use as many spectra as there are components present in the stellar system, obtained at different phase positions. This yields an analytically determined system of equations as long as noise (and further errors, e.g., from wavelength calibration) are negligible. Solving this set
of spectra results in decomposed spectra for each component. The result is commonly an average of solutions from several sets to reduce the noise of the decomposed spectra. Disentangling on the contrary, uses a larger set of observations that are spread over the whole
orbit of the system, and identifies moving contributions of each component in the composite spectra. Its strength is derived from an overdetermined situation, which makes noise a less degrading factor. As long as all data are of equal quality, the result will have a significantly higher signal-to-noise ratio (S/N). Furthermore, disentangling does not assume information about the radial velocities since the result is coupled to an optimisation on the orbital elements.

Spectral disentangling can also be defined as template-independent separation of different contributions of sources to an observed composite spectrum. These contributions are mainly the components of multiple stellar systems.
However, telluric lines and time-dependent line variability \citep[e.g.][]{2004A&A...422.1013H, 2009A&A...507..397H} can also be extracted from composite spectra. The first application of a decomposition procedure was
published by \citet{1991ApJ...376..266B}, where they used a tomographic method to separate the spectra of the binary AO~Cas. \citet{1994A&A...281..286S} published the disentangling method based on singular value decomposition (SVD).
They also implemented a global optimisation \citep[Downhill-Simplex,][]{Nelder01011965} method to find the best orbital fits to their observations. Independently, \citet{1995A&AS..114..393H} developed his famous code based on a Fourier transform of the composed spectra. Further development resulted in \texttt{KOREL} \citep[e.g.,][]{1997A&AS..122..581H}, \texttt{FDBinary} \citep[and fd3, see][]{2004ASPC..318..111I} and \texttt{CRES} \citep[see][]{2004ASPC..318..107I}. \texttt{CRES} uses SVD but it does not optimise the radial velocities of the observations.

A comparison between disentangling in Fourier and $\lambda$-space is given by \citet{2002FizBe..10..357I}. The disentangling in the $\lambda$-space is more flexible in the sense that (1) each spectra can have its own sampling, (2) each data point
can have its own weight, and (3) the output can be extended owing to the radial velocities (RVs) of the observations. In Fourier-disentangling all spectra need to have equal binning, weight and the output is a periodic function of the wavelength.
Our code is written to elevate point (3). Hence, the solution will cover an extended wavelength range, defined by the radial velocities of the observations. Point (2) could in both Fourier and $\lambda$-space cases be achieved
by rebinning all spectra to the same grid. However, it may be better to use different sampled data without resampling them to a common grid.

In Sect.~\ref{sec:method} we collect all necessary basics of spectral disentangling in the $\lambda$-space and describe the procedure. We present tests of the code with artificial data in Sect.~\ref{sec:tests}. An application to observations is presented in Sect.~\ref{sec:application} and the conclusion follows in Sect.~\ref{sec:conclusio}.

\section{Method of disentangling in ${\lambda}$ -space\label{}}
\label{sec:method}

\subsection{Formulation of the problem}
\label{sec:problem}

The time-series of spectra, spread over the orbital phase, needs to be carefully normalised. To account for the relative flux ratio of the components, additional data are required (e.g., from photometry).
We also note, that disentangling is only applicable to well-detached systems. However, the components can have different spectral types.

At first, we assume that the flux is constant with phase, i.e., the spectra from a possible eclipse are not used. The spectra are furthermore corrected to the heliocentric system and rebinned to a logarithmic wavelength scale
\begin{equation}
\label{Eq1}
 p=A\log(\lambda)+B,
\end{equation}
where $A$ and $B$ are constant and $p$ is the pixel number of the new scale. A given shift of the spectra by the Doppler effect yields a constant shift of $p$. If we assume that the resolution of the observations is sufficiently
high (in a first approximation given by the rotational velocity of the components) we can apply a linear transformation between observations and individual spectra. The solution vector
\begin{equation}
\label{Eq2}
 \overrightarrow{x}=(\overrightarrow{x_{1}}, \cdots, \overrightarrow{x_{k}})^{t}
\end{equation}
is given by the individual spectra of the $k$ components of the multiple system. If we collect all $n$ observations in
\begin{equation}
\label{Eq3}
 \overrightarrow{o}=(\overrightarrow{o_{1}},..., \overrightarrow{o_{n}})^{t}
\end{equation}
the problem to be solved is given by
\begin{equation}
\label{Eq4}
 \underline{M} \cdot \overrightarrow{x}=\overrightarrow{o}
\end{equation}
where $\underline{M}$ is the transformation matrix containing the radial velocity information of all spectra. This matrix can be further separated to
\begin{equation}
\label{Eq5}
 \underline{M}=
	      \begin{pmatrix}
	       \underline{N}_{1,1} & \cdots & \underline{N}_{k,1} \\
	       \vdots & \ddots & \vdots \\
	       \underline{N}_{1,n} & \cdots & \underline{N}_{k,n} \\
	      \end{pmatrix},
\end{equation}
where each of the sets $\underline{N}_{1,\jmath}, \dots, \underline{N}_{k,\jmath}$ images the component spectra to one observation and contains the RV information
\begin{equation}
\label{Eq6}
 \underline{N}_{\imath,\jmath}=\bordermatrix{~ & v_{\imath,\jmath} & ~ & ~ & ~\cr
		       ~ &\overbrace{0 \cdots 1} & \cdots & 0\cr
		       ~ & \vdots &\ddots & \vdots \cr
		       ~ & 0 & \cdots & 1 \cr},
\end{equation}
where $v_{\imath,\jmath}$ is the velocity in observation $\jmath$ of component $\imath$. As long as the SVD is applied to rank deficient and overdetermined systems, the solution behaves as a least square fit, i.e., it gives the
solution of smallest residuum, which is
\begin{equation}
\label{Eq7}
 r=\| \underline{M} \overrightarrow{x}-\overrightarrow{o} \|.
\end{equation}
This residuum can be used as the value to be minimised during the optimisation procedure. However, \citet{1994A&A...281..286S} optimised on the orbit fit to the RVs from the optimisation algorithm. To fulfil the criterion
of an overdetermined system we need at least one more observation since components, $n_{min}=k+1$, are to be disentangled at different phases of the system.

If static lines, e.g., tellurics, are present, we add an additional column of matrices, $\underline{N}_{k+1,\jmath}$, without shifting the diagonal. Since telluric lines vary in strength from observation to observation, we need to take this into
account. This can be done by using the relative line-depths as the elements for this matrix. Furthermore, heliocentric correction should not be applied since the static matrix does not account for shifts. In this case, the 
velocities are given by $v_{\imath, \jmath}=RV_{\imath, \jmath}+h_{\jmath}$ the sum of the radial velocities and the heliocentric correction. We note, that the minimum number of necessary observations $n_{min}=k+2$ increases by one.

\subsection{The procedure}
\label{sec:procedure}

\begin{figure}[h!]
 \begin{center}
   \begin{subfigure}{.5\textwidth}
  \centering
  \caption{}
  \includegraphics[trim=0cm 0cm 0.3cm 0cm, clip=true, width=85mm]{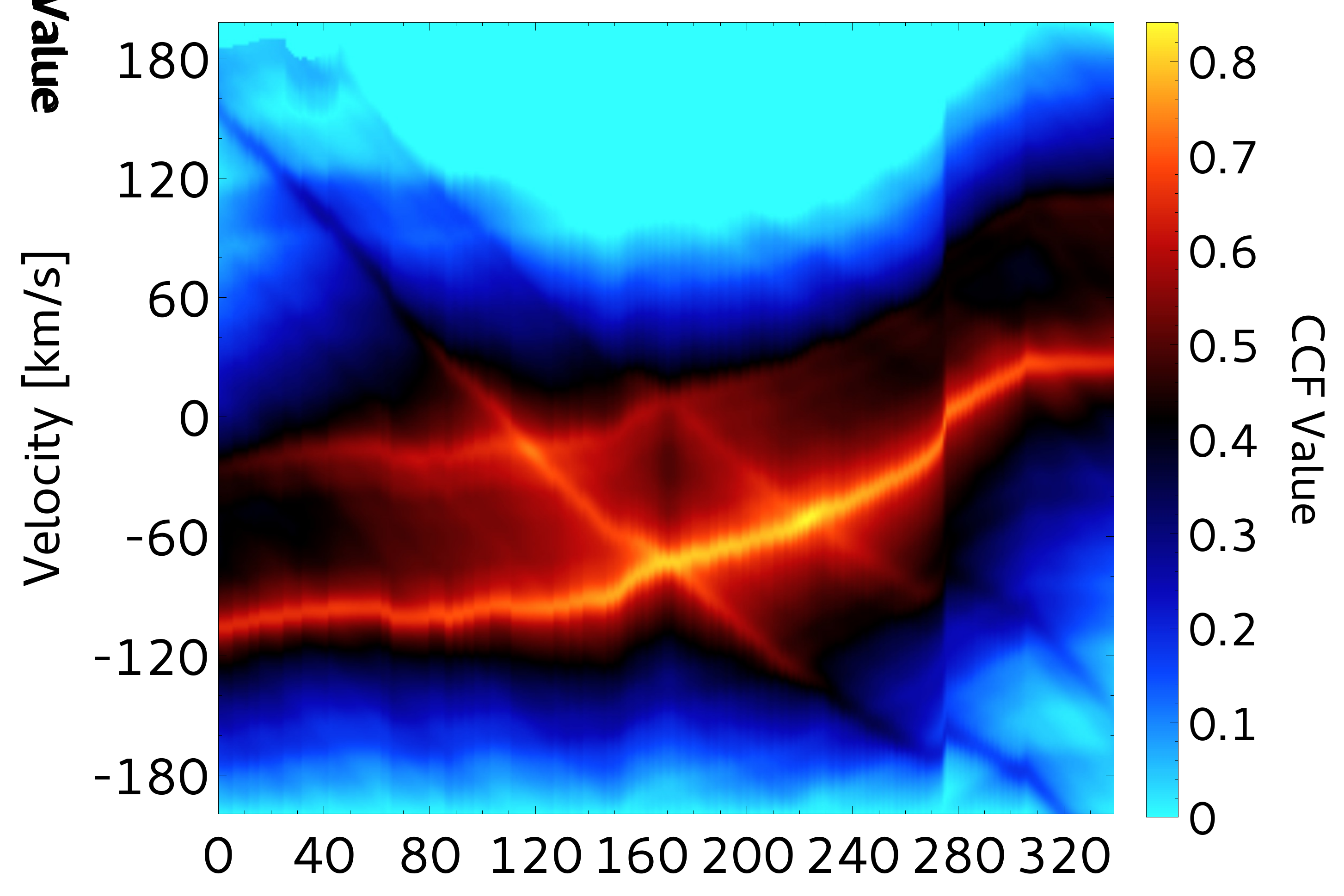}
  \label{Fig1:sfig1}
 \end{subfigure}
\begin{subfigure}{.5\textwidth}
\centering
\caption{}
 \includegraphics[trim=0cm 0cm 0.3cm 0cm, clip=true, width=85mm]{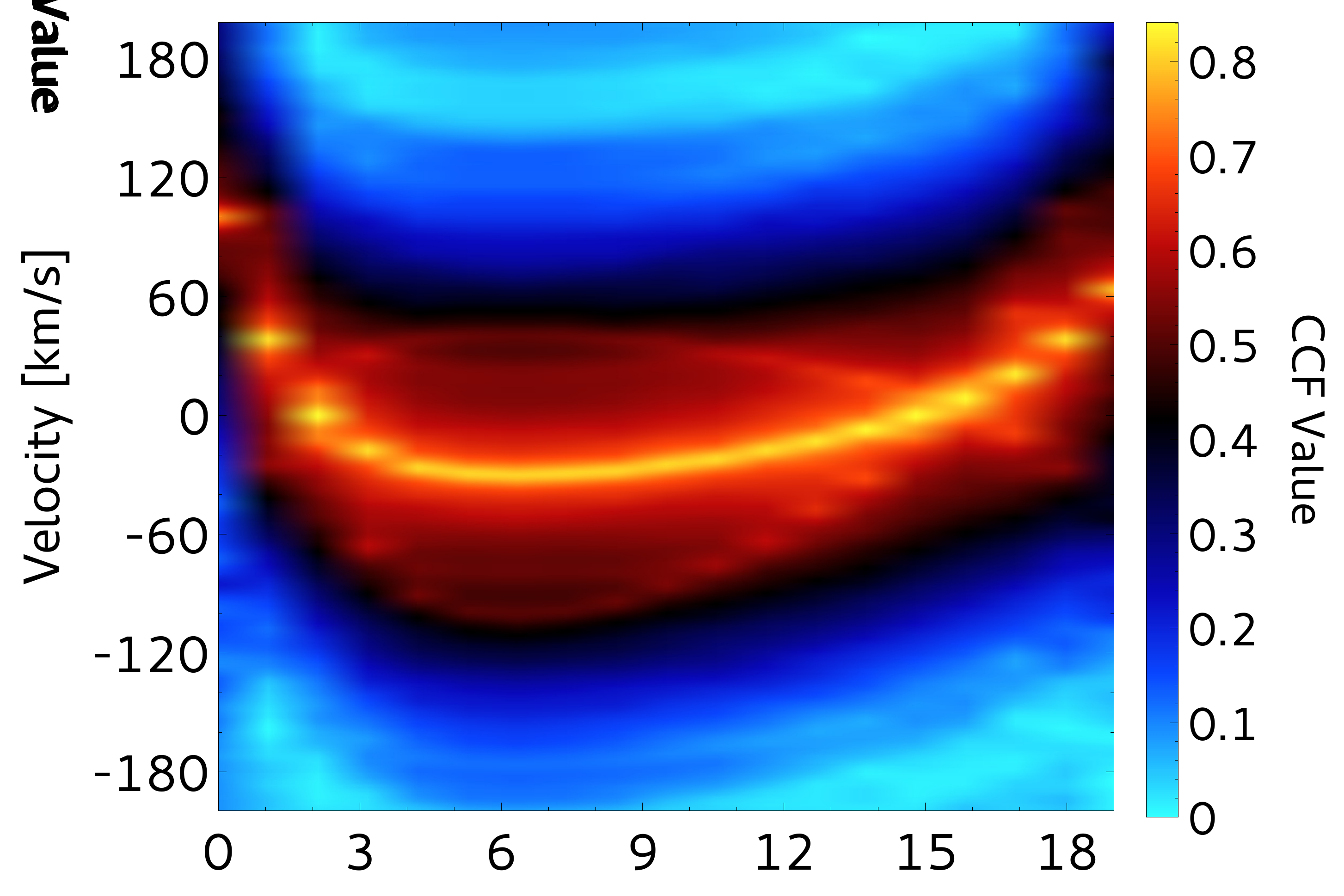}
  \label{Fig1:sfig2}
\end{subfigure} 
\end{center}
 \caption{Example of the 2D cross-correlation as provided by \textit{CroCo}.
 (a): All cross-correlations (340) performed on one spectrum;
 (b): The cross-correlations with maximum peak value for all (20) spectra equidistantly spread in phase of our artificial data.}
  \label{Fig1}
  \end{figure}

We assume that the observations are carefully normalised and corrected for the heliocentric velocity, which should be applied by the spectroscopic data reduction. However, heliocentric correction is not necessary if
the orbital motion is not of interest or if the velocities will be corrected afterwards. Moreover, in regions where telluric lines are present, no heliocentric correction should be applied. We have created an additional C++ GUI\footnote{We use the Qt library for GUI programming: www.qt.io} application (\texttt{CroCo}) for logarithmic rebinning 
and 2-dimensional cross-correlation, i.e., this tool is used to prepare the input data. The data are resampled to a new grid in the logarithmic scale. The step size $I$ is set by the user in parts of the smallest $\delta =\mathrm{min}(\log(\lambda_{\ell+1})-\log(\lambda_{\ell}))$ difference between the
sampling points on the logarithmic scale, such that the new sampling is $\delta^{\prime}=I \cdot \delta$. This sampling is also applied to the templates and all data are cropped to an identical wavelength region for the cross-correlation.
It is also possible to choose the spectral region of interest when it is not necessary to process the whole spectral range is not necessary. Linear interpolation is applied to find the values of the spectra on the new wavelength scale.
Furthermore, the continuum value is subtracted since we perform the cross-correlation for zero-normalised spectra. The cross-correlation sum for these data is

\begin{equation}
\label{Eq8}
 C(n)= {1 \over N \sigma_{s,t}} \displaystyle\sum_{m}^{N}(s(m)-\overline{s}(m))(t(m-n)-\overline{t}(m)),
\end{equation}
where $\overline{s} $ and $\overline{t}$ are the mean values and

\begin{equation}
\label{Eq9}
 \sigma_{s,t} \equiv \sigma_{s} \sigma_{t} = {1\over N-1} \sqrt{ \displaystyle\sum^{N}(s(m)-\overline{s}(m))^2(t(m-n)-\overline{t}(m))^2}
\end{equation}
the standard deviation product of the spectrum and template, respectively. Based on the new wavelength-sampling, \texttt{CroCo} performs the cross-correlation of all possible combinations of the templates in the specified velocity ranges with the observed spectrum. These ranges are defined by a value for maximum separation of the templates and maximum shift against the rest wavelength. In Fig.~\ref{Fig1}, we show an example of a cross-correlation for one spectrum (a) and the whole
set (b) around the \ion{Li}{i} $\lambda$6708{\AA} line of an artificial binary. 
All performed 2-dimensional cross-correlations are shown. The program picks out the one with the maximum value. In regions where no overlap between template and observation exists, we add 0 to the cross-correlation sum.
Cross-correlation was performed using the templates from which the spectra themselves were constructed. More details on these data follow in Sect. \ref{SB2}.

\begin{figure}[h!]
 \begin{center}
 \includegraphics[width=85mm]{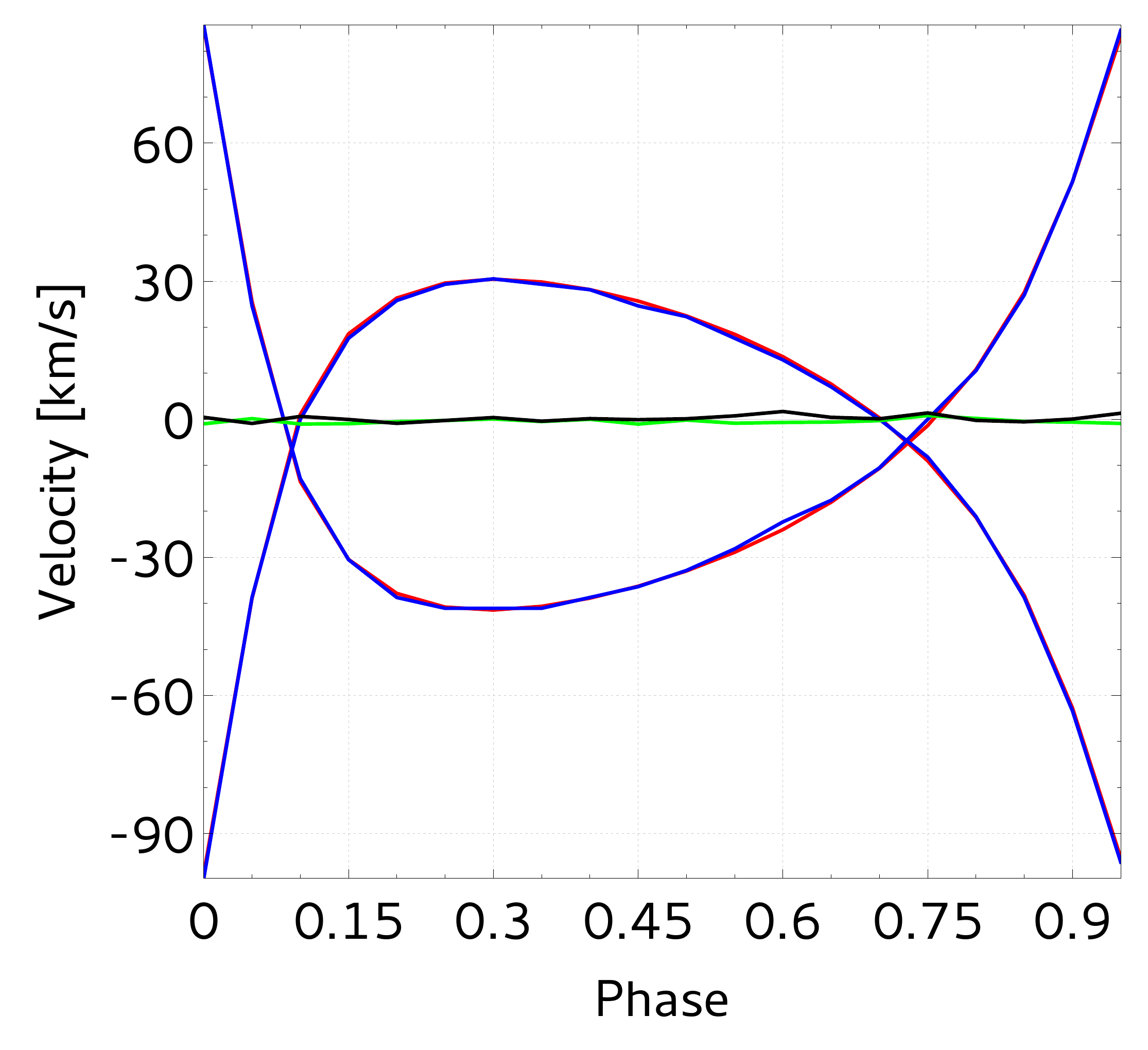}

\end{center}
 \caption{Orbit of our test system. Red is calculated and blue is the result from the cross-correlation. Green and black show the differences between calculated and cross-correlation for both components. The velocity sampling was about 1.17\,km/s and we have not fitted the CCF peak.}
  \label{Fig2}
  \end{figure}

For this artificial binary, we applied an orbit comparable to Mizar ($\zeta$~UMa) with an eccentricity of $e=0.537$, systemic velocity $\gamma=-5.6$\,km/s, and amplitudes of $K_{1}=68.6$\,km/s and $K_{2}=67.6$\,km/s for the primary and secondary, respectively. 
With the Binary Tool implemented in \texttt{CroCo}, two template-spectra can be combined to a composite spectra that covers the specified orbit. These data can be used to find systematic errors and to cross-check with the results of the observations. The orbit and the result from the cross-correlation are shown in Fig.~\ref{Fig2}. We note here, that we do not fit the correlation peak for a more precise determination because these values will be optimised during the disentangling procedure. However, the cross-correlation data are stored as ASCII files for further processing.
The velocities from the cross-correlation can be used for the disentangling when a preliminary orbit is not known. As we describe below, optimisation can be performed directly on the RVs of each component and each observation
or on the orbital parameters. Optimisation on the velocities can be advantageous in case of orbital distortions, e.g, a third component.

During the orbital motion, the wavelength window will vary according to the RV value.
This leads to slight differences of the wavelength range that each individual observation covers. To prevent edge errors and to gather the maximum wavelength information, we construct the matrices in Eq.~\ref{Eq6} such that they cover the maximum wavelength range. This in turn will enlarge the wavelength range of the solution vectors (Eq.~(\ref{Eq7})).
As described in the introduction, this would not be possible using a procedure that is based on a Fourier-transform.
  
The disentangling code \texttt{Spectangular} uses the \texttt{Armadillo} C++ linear algebra library \citep[see][]{Sanderson_10} for efficient matrix and vector operations. It also provides three different algorithms for the SVD, the classical full SVD, an economical solution (ECON), and
divide-and-conquer (DaC) method. For most data sizes, DaC is by far the fastest procedure and is implemented in the global optimisation algorithm. Nevertheless, for very small data sets, the classical SVD could be faster since it does not require any pre-calculations, e.g.,
DaC needs to separate the large transformation matrix $\underline{M}$ into smaller ones and treats them separately. Disentangling without any optimisation can also be performed with all three methods. Furthermore, we use the
\texttt{OpenBLAS} library, which provides parallel computing routines. This is especially advantageous for matrix operations, since it reduces the time for these calculations by approximately the reciprocal of the number of threads used.

The most time-consuming part of the optimisation algorithm is the initiation. In this first step, it creates the initial simplex which has $(kn+1)$ points, i.e., the SVD is performed $(kn+1)$-times for the whole data set.
This initial data is saved in case one wants to start the optimisation again. We note that the variation of the RV values to create this simplex is not random, i.e., for a given set of input RVs (or orbital parameters) the initial simplex will always be the same. This will be described in more detail in Sect. \ref{DSM}. After this initiation the algorithm performs as many iterations as specified by the user. Depending upon the data and initial values, the SVD is performed multiple times during one iteration. The most
time-intensive transformation of the simplex is a total contraction, where $kn$ points need to be re-calculated, i.e., it takes almost as long as the initiation itself.
We have implemented direct optimisation on the radial velocities and on the orbital elements. When optimised on orbital elements, the initiation takes only eight 
(seven orbit elements) evaluations and, in general, this procedure leads to faster convergence. The value to be minimised in both cases is the
residuum between result and observations given by Eq.~(\ref{Eq7}). Additionally, it is also implemented to optimise on the peak-deviation, i.e., the maximum deviation of a single spectral bin to the disentangling result. This criterion, however, is not recommended since it is very sensitive to remaining cosmics, pixel errors, etc.

The resulting optimisation data are saved again to a local file and can be used if and when the optimisation needs to be continued. If the residuum $r$ was reduced after a transformation, the result can be plotted in the main window
of the program as a live view to control the procedure. After each evaluation of the SVD, the optimisation can be aborted in case that $r$ is not minimsed any longer.
Currently, the code is implemented for SB1 and SB2 data and the user can choose whether telluric (static) lines are present or not.

\subsection{Details on downhill simplex implementation}
\label{DSM}

For the downhill simplex method, it is common to use the standard deviation of all $(kn+1)$ function values (the residuum of the SVD in our case) of the simplex as the abort criterion for the optimisation.
However, it is implemented to set the number of iterations of the downhill simplex algorithm (DSA). The arithmetic mean as well as the standard deviation of all residua of the simplex are calculated and written to the logfile after each iteration.

As already mentioned, the DSA performs transformations of the initial simplex. There is no general rule for constructing this initial simplex. In the case of optimisation on each RV, the initial simplex is
constructed as
\begin{equation}
 \label{Eq10}
 \overrightarrow{P_i} = \overrightarrow{P_0} + \delta RV \cdot \overrightarrow{e}, \ \ i = 1 \dots N,
\end{equation}
where $P_0$ is the point defined by the initial values, $\delta RV$ is the change, $\overrightarrow{e}$ is the unit vector, i.e., only one of
the values differs from point to point, and $N=kn$ is the number of variables. The change in RV can be set by the user in fractions of the binning $B_v$ of the spectra in velocity space, i.e., $\delta RV = S B_v$, where $S$
is the user-defined step size. In the case of optimisation on the orbital parameters, the step of each parameter is user-defined and $N=7$ for the number of orbital parameters of an SB2 system and with the relation $\omega_2 = \omega_1 + \pi$ for the longitude of periastron for the secondary.

Furthermore, the coefficients for the transformations can also be changed. There are four transformations: (1) reflection
\begin{equation} 
 \label{reflection}
 \overrightarrow{R} = \overrightarrow{Z} + \alpha (\overrightarrow{Z}-\overrightarrow{P}_w),
\end{equation}
of the worst point $P_w$ (corresponds to highest $r$) over the centroid $\overrightarrow{Z}$ of the simplex, where $\alpha=1$ is the reflection coefficient. (2) expansion
\begin{equation}
 \label{expansion}
 \overrightarrow{E} = \overrightarrow{Z} + \gamma (\overrightarrow{Z}-\overrightarrow{R}),
\end{equation}
over the reflected point, where $\gamma=2$ is the expansion coefficient. (3) single contraction
\begin{equation}
 \label{contraction}
 \overrightarrow{C} = \overrightarrow{Z} + \beta (\overrightarrow{P}_w-\overrightarrow{Z}),
\end{equation}
of the worst point towards the centroid, where $\beta=0.5$ is the contraction coefficient. (4) total contraction
\begin{equation}
 \label{totcont}
 \overrightarrow{P'}_i = \overrightarrow{P_b} + \beta_{tot}(\overrightarrow{P}_i - \overrightarrow{P_b}),\ i\neq b,
\end{equation}
towards the best point $\overrightarrow{P_b}$, where $\beta_{tot}=0.5$ is the total contraction coefficient. All given values for the coefficients are defaults and used if user values are not defined.

We note, that the DSA could stagnate and minimisation of the residuum is not achieved even after many iterations. In this case the optimisation should be aborted. The current 
best values for the parameters should be used as the new start values to start a new optimisation. Hence, an automatic re-initiation option is implemented so that optimisation
will be reinitiated after DSA has passed two iterations without a change in mean and STD values of the simplex points.
Another complication can occur in computing the correct residuum between solution and observations at the edges of the spectra. The solution can lose its overdetermination, i.e., the solution is poorly defined at the edges.
In this region the entries of the solution vector $\overrightarrow{x}(edge)$ (more precisely: the corresponding singular values) do only represent rounding errors. Computing the residuum against the whole spectra would yield wrong (higher) values of the residuum. Hence, there is the option
to calculate the residuum in a (smaller) user-defined wavelength range.

\subsection{Error estimation}
The common way \citep[e.g.][]{doi:10.1021/ac00159a002, Nelder01011965} is to compute the curvature matrix
\begin{equation}
 \label{Curv}
 \Gamma_{\imath \jmath} = \frac{\partial^2 \chi^2}{\partial p_\imath \partial p_\jmath},
\end{equation}
where $\chi^2 = \sum w_\imath(y_\imath-f_\imath)^2$. From this matrix one obtains the variance-covariance matrix according to
\begin{equation}
 \label{VarCo}
 \epsilon_{\imath \jmath} = s^2(\Gamma^{-1})_{\imath \jmath},
\end{equation}
where $s^2 = \chi^2/(n-N)$. When the off-diagonal elements of $\underline{\epsilon}$ can be neglected, the standard deviation of parameter $\imath$
is given by $\sigma_{p_i}=\sqrt{\epsilon_{\imath \imath}}$. To fit a quadratic function to the error surface additional points $\overrightarrow{P}_{\imath \jmath} = (\overrightarrow{P}_\imath + \overrightarrow{P}_\jmath)/2,\ \imath \neq \jmath$ are required.
This leads to an additional $(N+1)N/2$ necessary SVDs, i.e., Eq. (\ref{Eq4}) needs to be solved 28$\times$ in case of an SB2 system to estimate the error. It becomes computationally expensive for large data sets and for optimisation on the individual RVs, e.g, 10 spectra of an SB2 system would require 210 SVDs of Eq. (\ref{Eq4}).
However, it is an acceptable and necessary effort for the orbital parameters and the advantage is that no numerical increments for the derivatives need to be assumed (Eq.(\ref{Curv})).

Nevertheless, there are two cases in which the quadratic fit is not applicable. Firstly, the final simplex could be too small, i.e., the error estimation is dominated by rounding errors.
Secondly, the opposite case where the final simplex is too large for a quadratic fit. As pointed out by \cite{doi:10.1021/ac00159a002}, the former case is the more important one. Indeed, if the
optimisation is performed with already good parameters, it is possible that one of them is only changed by a small value or, even worse, not changed at all. This will
make it impossible to compute the curvature matrix, since inversions are involved and they would be singular. In this case the critical parameter is varied in one vertex of the simplex and the residuum recalculated.
However, this decision requires user-input and one should be aware that there is also the possibility of creating new optimisation data by using the best parameters and starting the optimisation again only for a few iterations.

Finally, based on experience, we note that a good alternative is to perform the optimisation on individual RVs and apply
an orbit fit to the optimised RVs. This provides errors for the orbital parameters. Furthermore, errors introduced by a third component and 
uncertainties on the wavelength axis of the spectra (e.g., due to calibration errors) will have less of an effect on the resulting spectra.

\section{Test with artificial data}
\label{sec:tests}

\subsection{Performance}

\begin{figure}[h!]
 \begin{center}
  \includegraphics[trim=0cm 0cm 0.1cm 0cm, clip=true, width=85mm]{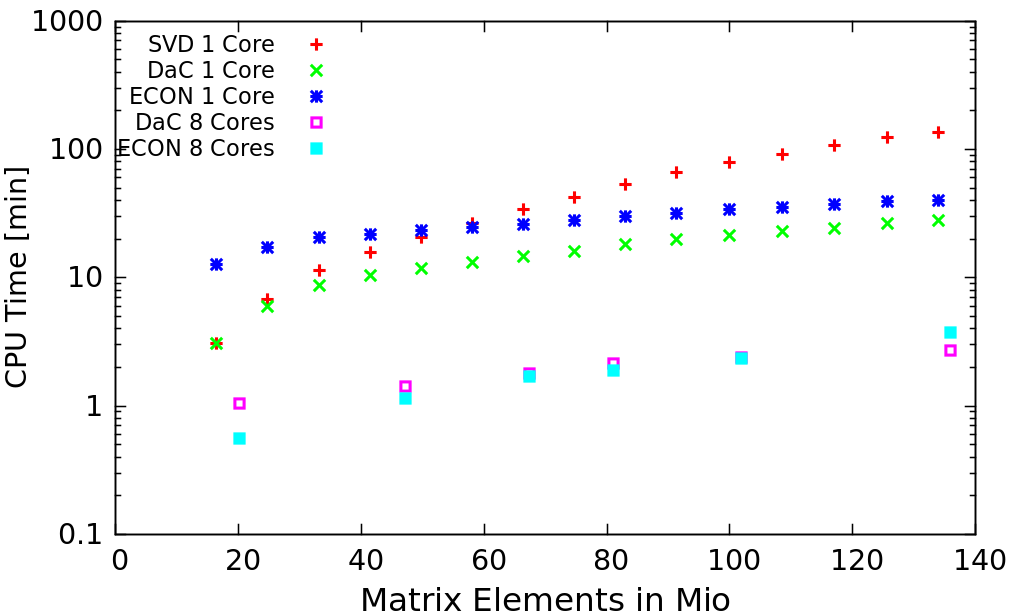}
\end{center}
 \caption{CPU time vs. data size for the three SVD methods for computation on one and eight cores. SVD is the standard full SVD, ECON the economical SVD and DaC is for SVD with devide-and-conquer method.\\}
  \label{Fig3}
  \end{figure}

We used artificial data with different sampling to test the computation time, depending on data size. In Fig.~\ref{Fig3}, we show a comparison between single core and eight logical cores. The spectra had a length of around 2,000\,points after rebinning to the logarithmic wavelength scale. Hence, the different data size comes from the number of used spectra. The time depends on both number of rows $u$ and columns $v$ of the $u \times v$ transformation matrix. 
For a rectangular matrix, $ v < u $, it is sufficient to calculate only $v$ singular values. The singular matrix is then a quadratic $v \times v$ matrix and only these column vectors need to be considered (this is
one of three economical SVD methods - thin SVD).

We note that our code development workstation had four physical cores and we enabled hyperthreading to split these into eight logical cores. Using eight threads, the computing time is reduced by a factor of eight.

\begin{figure}[h!]
 \begin{center}
  \includegraphics[trim=0cm 0cm 0.1cm 0cm, clip=true, width=85mm]{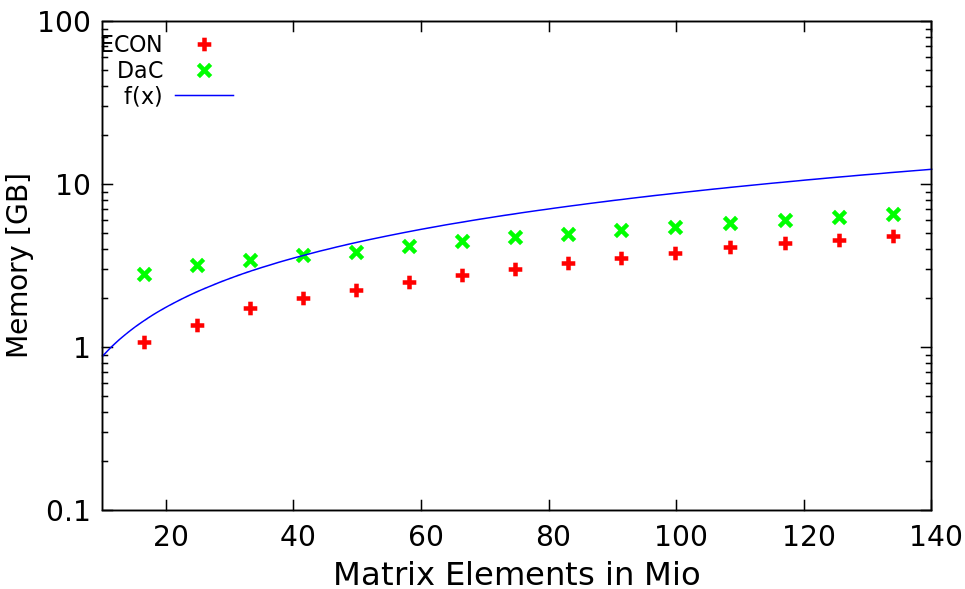}
\end{center}
 \caption{Memory vs. data size for the economical (ECON) SVD, SVD with divide-and-conquer method (DaC) and the memory estimation $f(n)$ (see text).}
  \label{Fig4}
  \end{figure}

The memory consumption can be estimated by the number of elements of the transformation matrix $\underline{M}$. Let $l_{p}$ be the number of spectral bins on the logarithmic scale. Neglecting the additional columns that are due to the 
velocity shift we get $v=2l_{p}$ and $u=nl_{p}$ for the number of columns and rows, respectively. SVD decomposes $\underline{M}$ into three matrices of dimensions $v \times v$, $u \times v$ and $u \times u$, respectively. 
We therefore get  the following relation for memory usage: $\mathrm f(n)=l_{p}^{2}(n(n+2)+4)\cdot8\cdot10^{-9} [GB]$. We show this dependency together with collected data for ECON and DaC in Fig. \ref{Fig4}. 

\subsection{SB1 and static lines}

\begin{figure}[h!]
 \begin{center}
 \includegraphics[width=85mm]{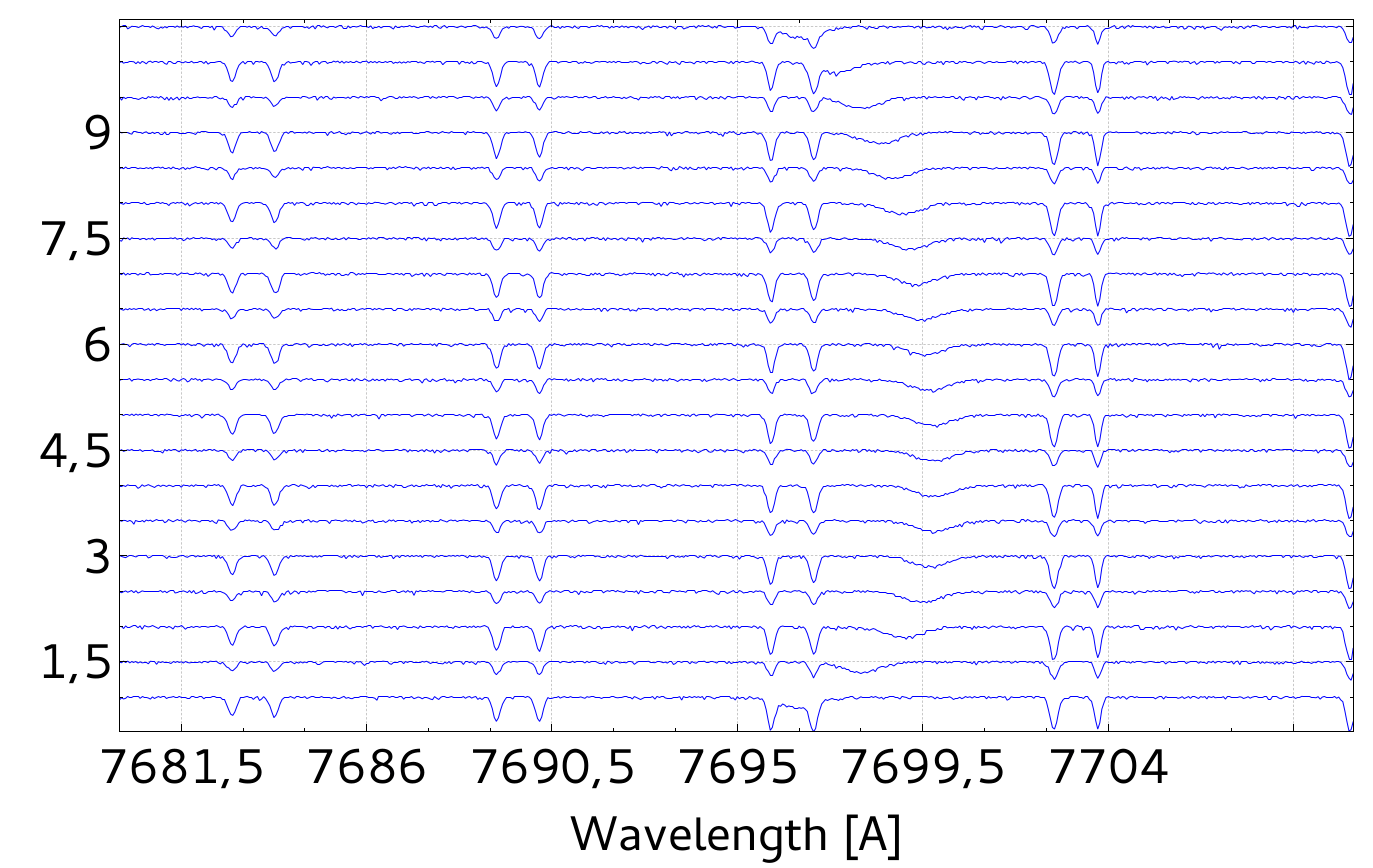}
\end{center}
 \caption{Artificial time-series of our SB1 system with the stellar \ion{K}{i} $\lambda$7699{\AA} line in the variable A band of O\textsubscript{2}.}
  \label{Fig5}
  \end{figure}

\begin{figure}[h!]
 \begin{center}
   \begin{subfigure}{.5\textwidth}
  \centering
  \caption{}
  \includegraphics[width=85mm]{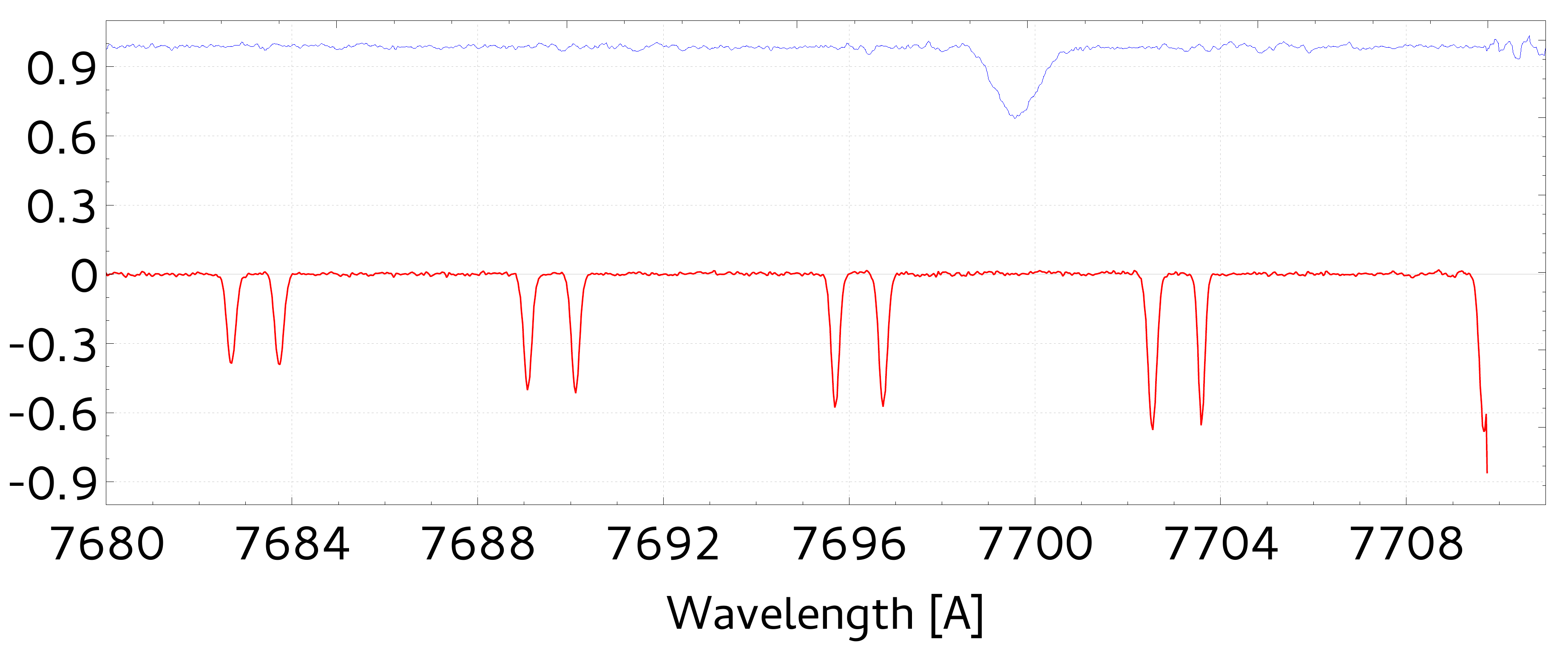}
  \label{Fig6:sfig1}
 \end{subfigure}
\begin{subfigure}{.5\textwidth}
\centering
\caption{}
 \includegraphics[width=85mm]{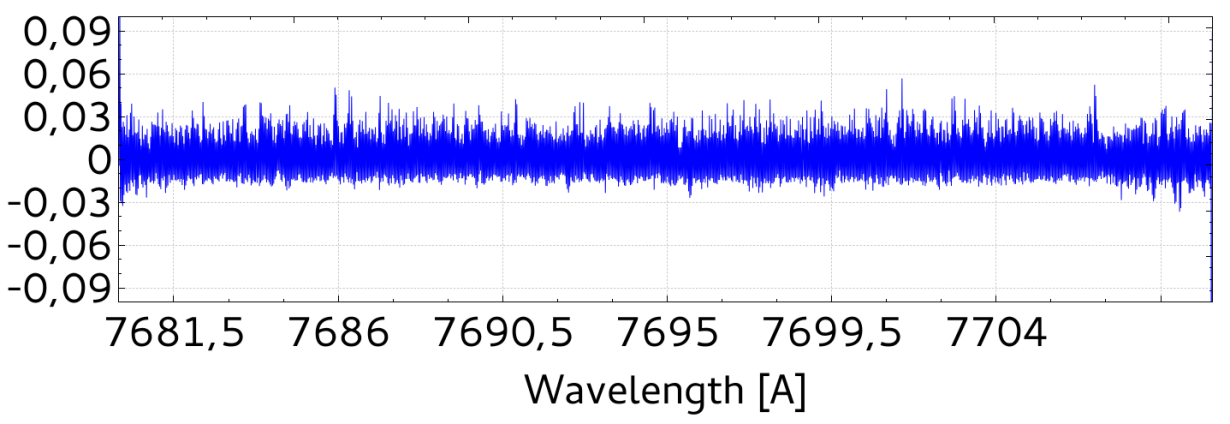}
  \label{Fig6:sfig2}
\end{subfigure}
\begin{subfigure}{.5\textwidth}
\centering
\caption{}
 \includegraphics[width=85mm]{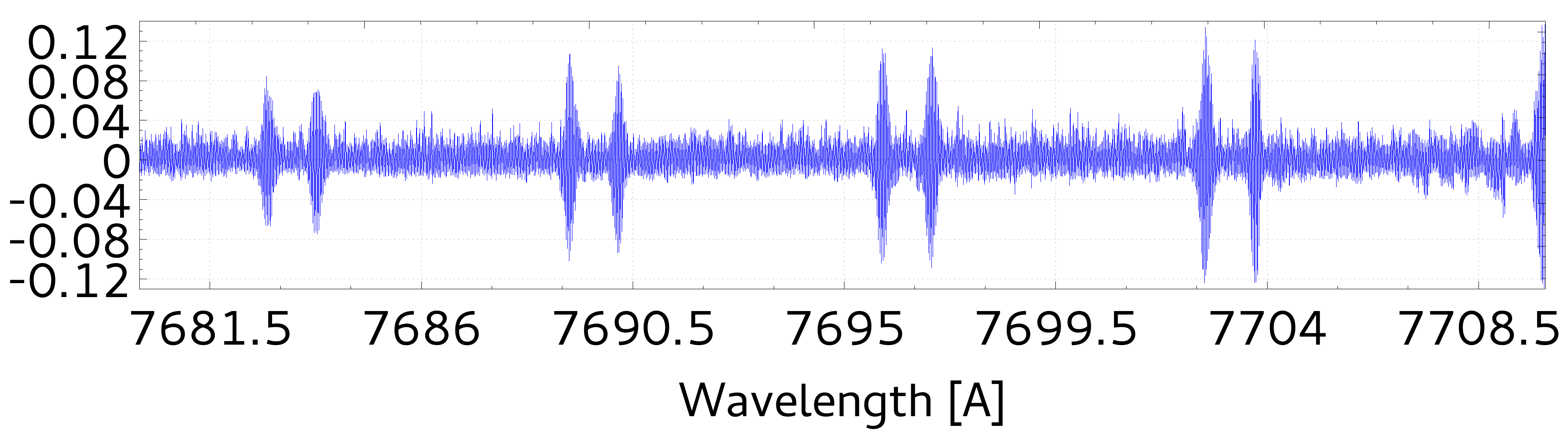}
  \label{Fig6:sfig3}
\end{subfigure} 
\end{center}
 \caption{Result of the disentangling with optimisation of our SB1 system with the stellar \ion{K}{i} $\lambda$7699{\AA} line in the variable A band of O${_2}$.
 (a): Absorptions from the SB1 line and variable telluric lines; (b): Differences between all spectra and result; (c): Differences between all spectra and result without accounting for line-strength variation.} 
  \label{Fig6}
  \end{figure}

Here we demonstrate the disentangling of lines of SB1 systems that are blended, e.g., from telluric features. This is the case for the \ion{K}{i} $\lambda$7699\,{\AA} line, e.g., in $\epsilon$~Aurigae \citep{2014AN....335..904S},
which is hidden in the A-band of O$_{2}$ around $\lambda$7600\,{\AA}. To demonstrate this, we created artificial data with an absorption at the \ion{K}{i} line wavelength and applied the orbital parameters, as described
in Sect.~\ref{sec:procedure}, and the time-series, as shown in Fig.~\ref{Fig5}. To get more realistic data, we added Gaussian noise resulting in an S/N of 100. Furthermore, we varied the line-strength of the telluric features and assume that the ratios
are measured and known. Hence, the matrix $\underline{N}_{k+1,\jmath}$ is filled with these ratios. We alternately changed the line-strength from spectrum to spectrum by a factor of 0.5.
Disentangling was performed with three iterations of the optimisation algorithm. The result of the disentangling is shown in Fig.~\ref{Fig6}. The top panel in Fig.~\ref{Fig6:sfig1} shows the disentangled SB1 line and the telluric lines. Fig.~\ref{Fig6:sfig2}, at the bottom, shows the differences between the disentangled result and the composite spectra. These differences are computed with all composite spectra according to $\underline{M}\overrightarrow{x}-\overrightarrow{o}$. We note, that the telluric features would also be successfully disentangled if the ratios are not given. In this case, the variation of line-strength would be
present in the differences between observations and result. This is shown in the bottom panel of Fig.~\ref{Fig6}

As described previously, the result vectors (\ref{Eq2}) cover a slightly larger wavelength range owing to the orbital motion. This can also be seen in  Fig. \ref{Fig6:sfig1} between the telluric and the stellar spectrum.
  
\begin{figure}[h!]
 \begin{center}
   \begin{subfigure}{.5\textwidth}
  \centering
  \caption{}
  \includegraphics[width=85mm]{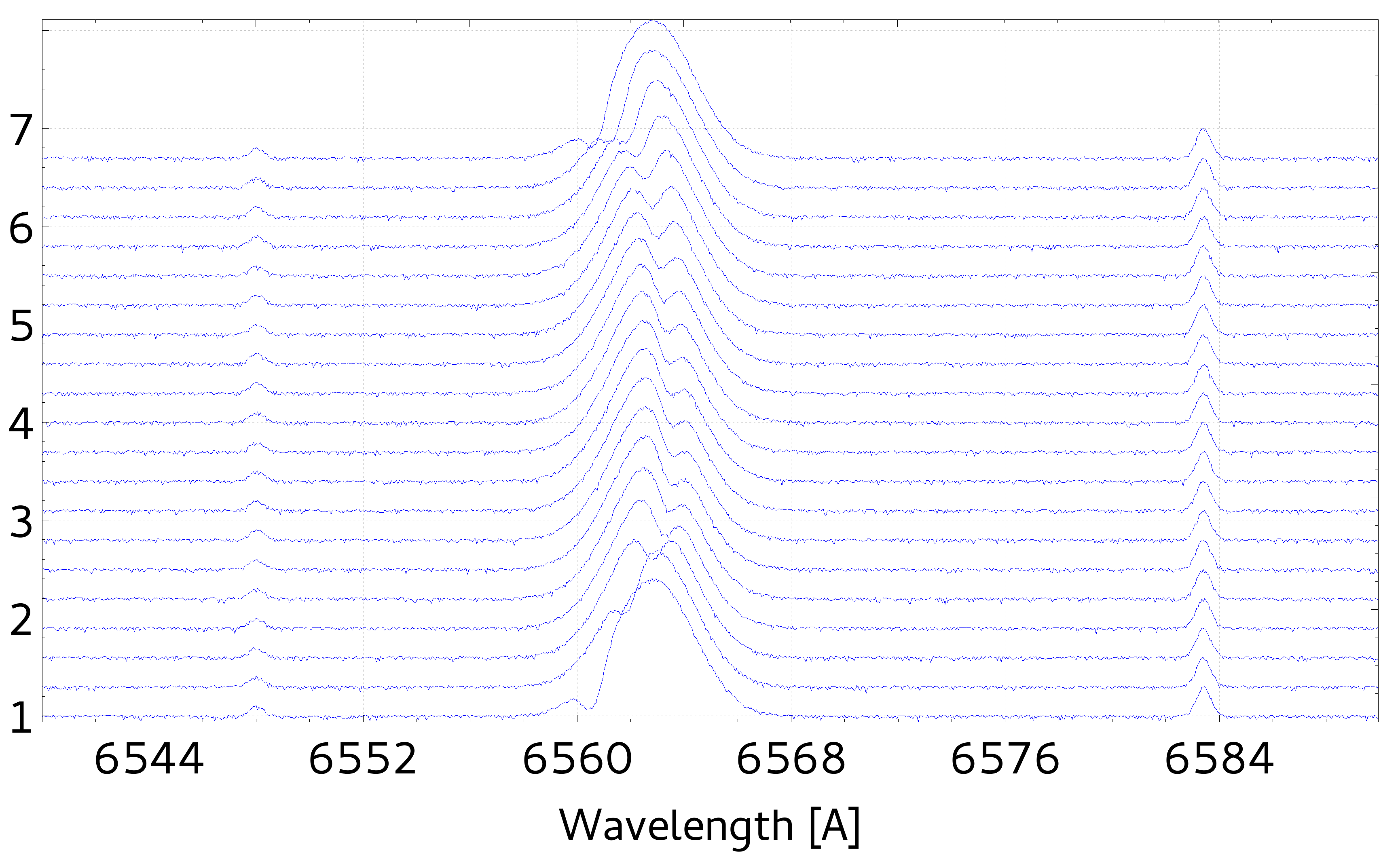}
  \label{Fig7:sfig1}
 \end{subfigure}
\begin{subfigure}{.5\textwidth}
\centering
\caption{}
 \includegraphics[width=85mm]{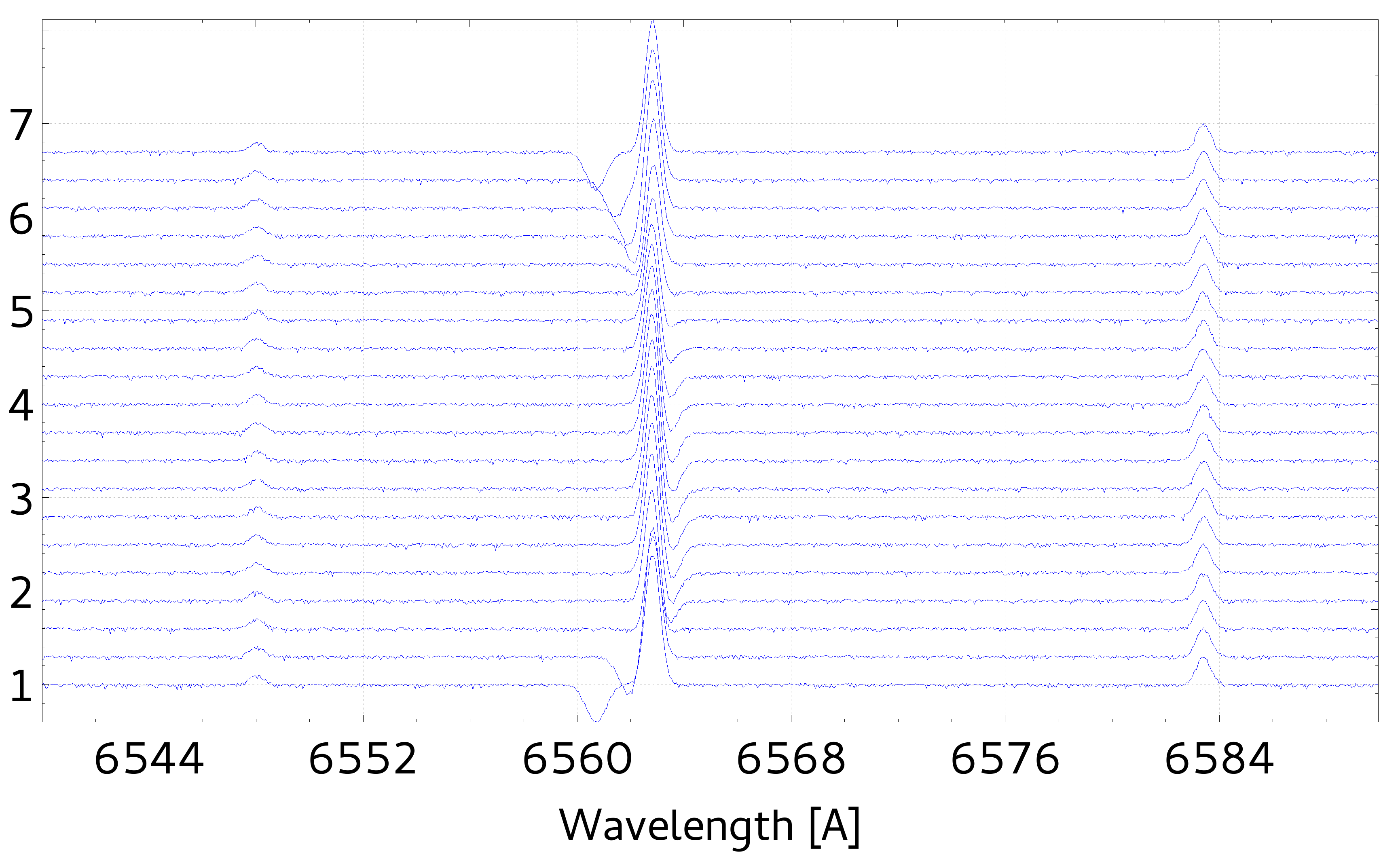}
  \label{Fig7:sfig2}
\end{subfigure} 
\end{center}
 \caption{Artificial time-series of our SB1 system with the superimposed broad (a) and narrow (b) emission.}
  \label{Fig7}
  \end{figure}

Additionally, we generated spectra with superimposed emission lines to simulate stars embedded in emission nebulae. We used emission line data
from \citet{1992ApJ...389..305O} of M42 and added Gaussian noise to our data. Again, we used the same orbit parameters as described in Sect.~\ref{sec:procedure} to model an absorption that crosses a static emission. We show the time-series of these data in Fig.~\ref{Fig7}, where we created two sets of data. These sets differ in the width of the emission-line compared to the width of the absorption-line.

\begin{figure}[h!]
 \begin{center}
 \begin{subfigure}{.5\textwidth}
  \centering
  \caption{Static emission (red) and SB1 absorption (blue).}
  \includegraphics[width=85mm]{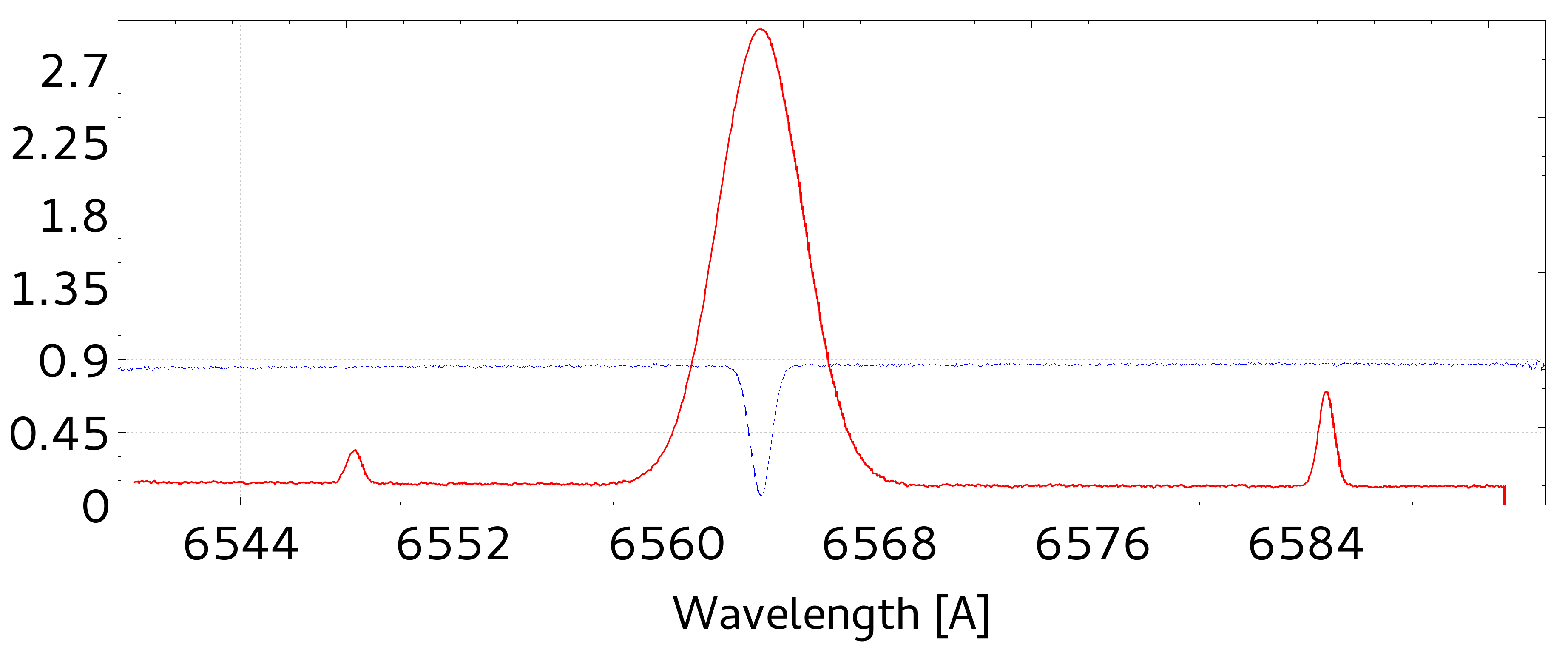}
  \label{Fig8:sfig1}
 \end{subfigure}
\begin{subfigure}{.5\textwidth}
\centering
\caption{Differences}
 \includegraphics[width=85mm, height=25mm]{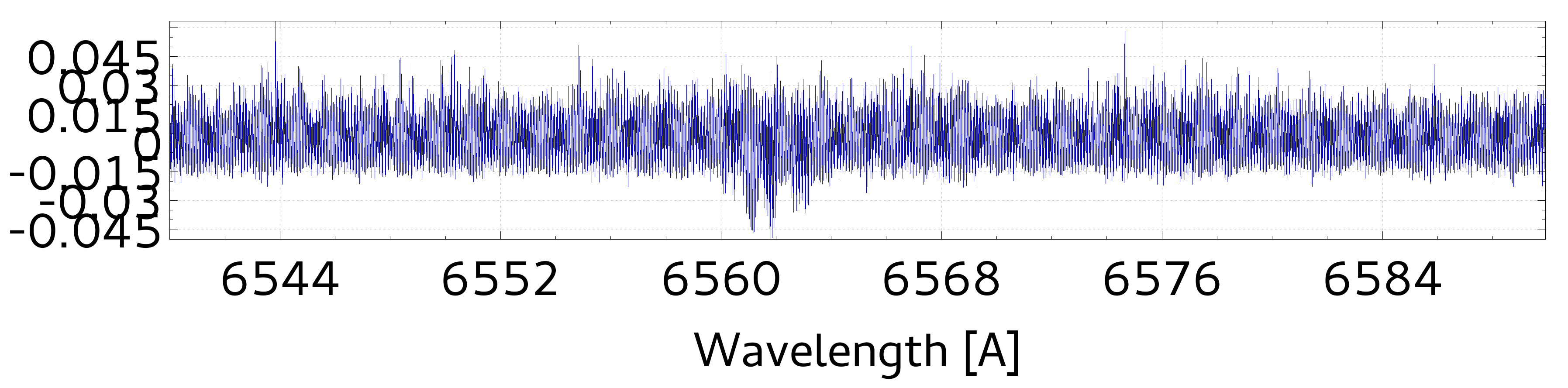}
  \label{Fig8:sfig2}
\end{subfigure} 
\end{center}
 \caption{Disentangling result of the superimposed broad emission-line of the data shown in Fig. \ref{Fig7:sfig1}.}
  \label{Fig8}
  \end{figure}

\begin{figure}[h!]
 \begin{center}
  \begin{subfigure}{.5\textwidth}
  \centering
  \caption{Static emission (red) and SB1 absorption (blue).}
  \includegraphics[width=85mm]{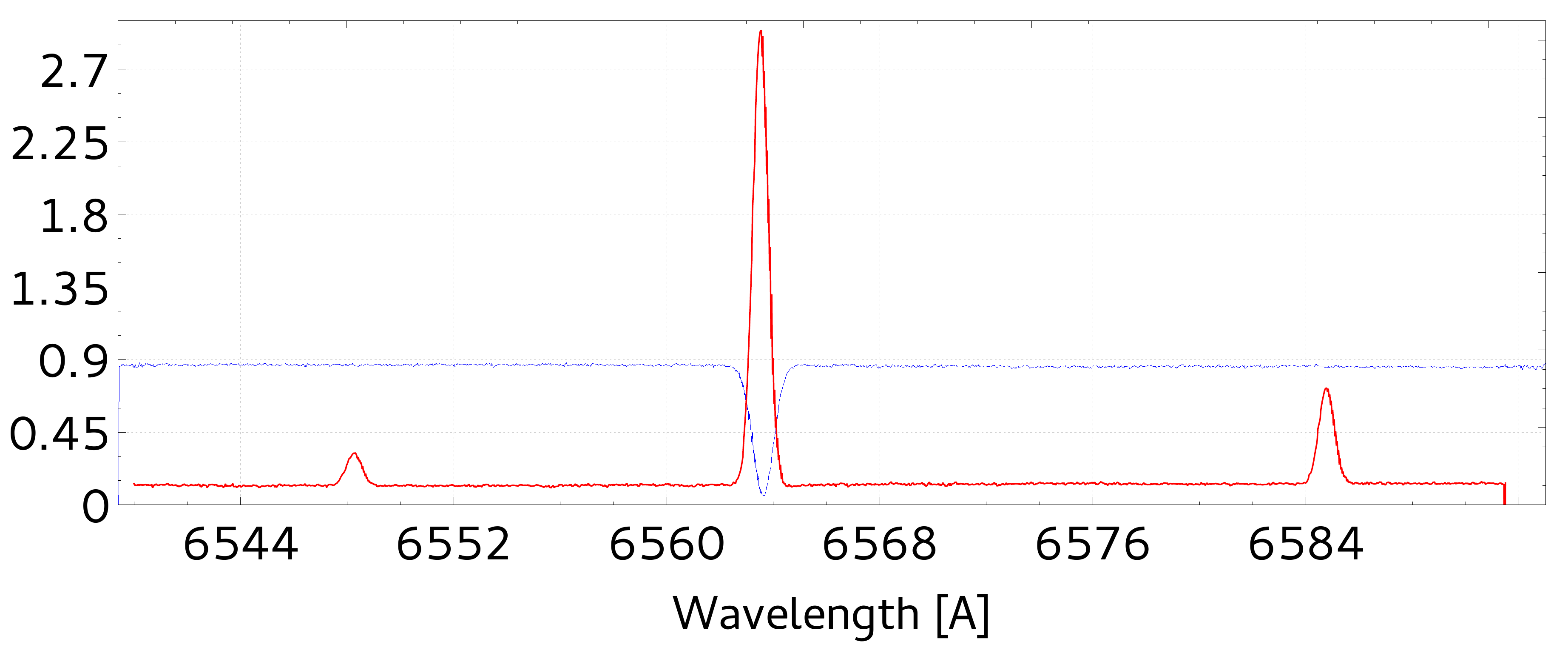}
  \label{Fig9:sfig1}
 \end{subfigure}
\begin{subfigure}{.5\textwidth}
\centering
\caption{Differences}
 \includegraphics[width=85mm]{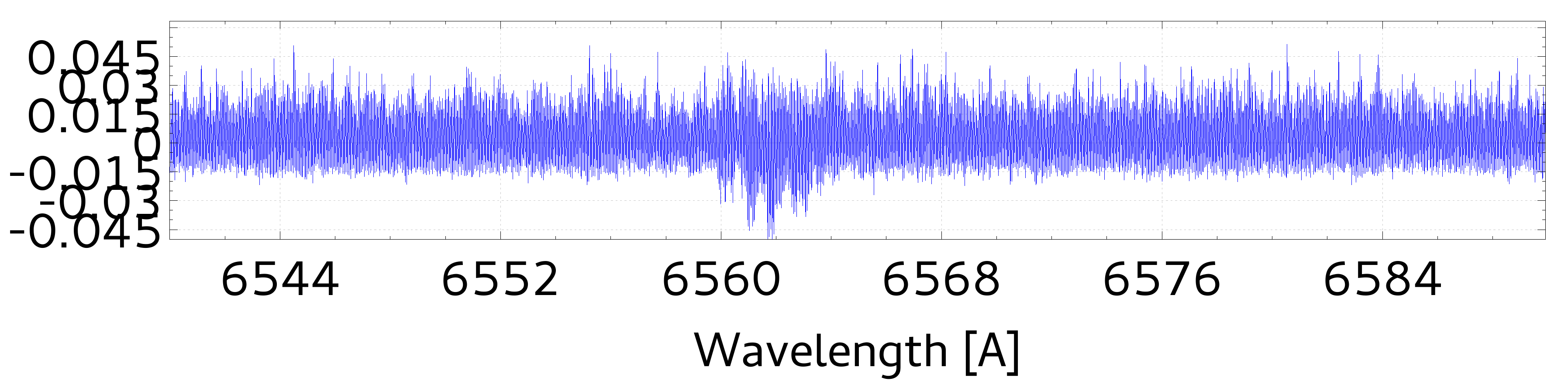}
  \label{Fig9:sfig2}
\end{subfigure} 
\end{center}
 \caption{Disentangling result of the superimposed narrow emission-line of the data shown on bottom in Fig. \ref{Fig7:sfig2}.}
  \label{Fig9}
  \end{figure}

The disentangling results after three iterations of optimisation are shown in Fig.~\ref{Fig8} for the broad emission and in Fig.~\ref{Fig9} for the narrow emission-line, respectively. The wavelength range of the result for the absorption spectra is, as expected, again slightly larger than that for the static lines. The residuum is of the order of 4.8 $\cdot$ 10$^{-5}$ in both cases.
  
\subsection{Artificial SB2 system}
\label{SB2}

\begin{figure}[h!]
 \begin{center}
 \includegraphics[width=85mm]{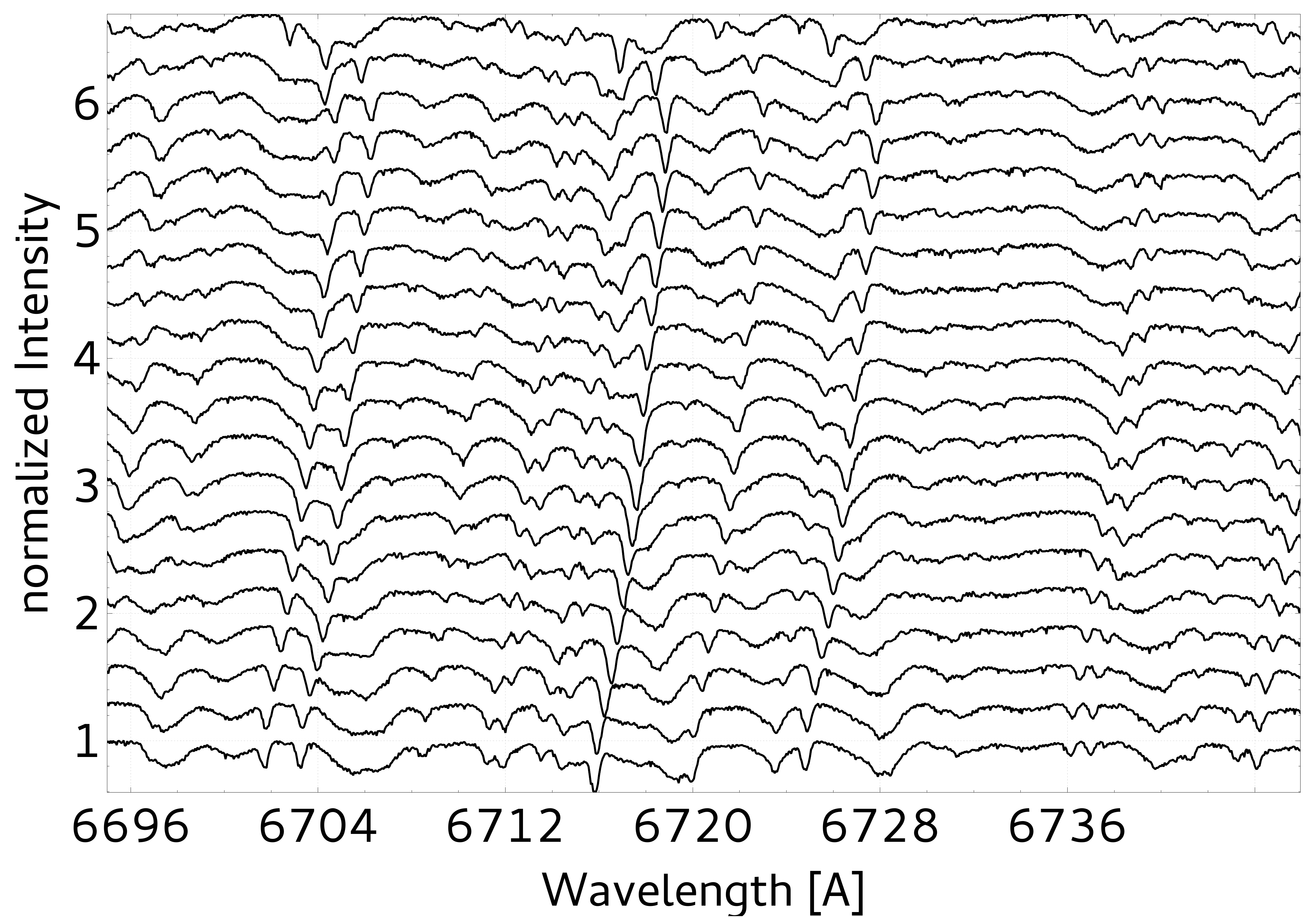}
\end{center}
 \caption{All 20 spectra of our artificial binary around the \ion{Li}{i} $\lambda$6708{\AA} line with S/N of 100.}
  \label{Fig10}
  \end{figure}

\begin{figure}[h!]
 \begin{center}
  \begin{subfigure}{.5\textwidth}
  \centering
  \caption{Component spectra}
  \includegraphics[width=85mm]{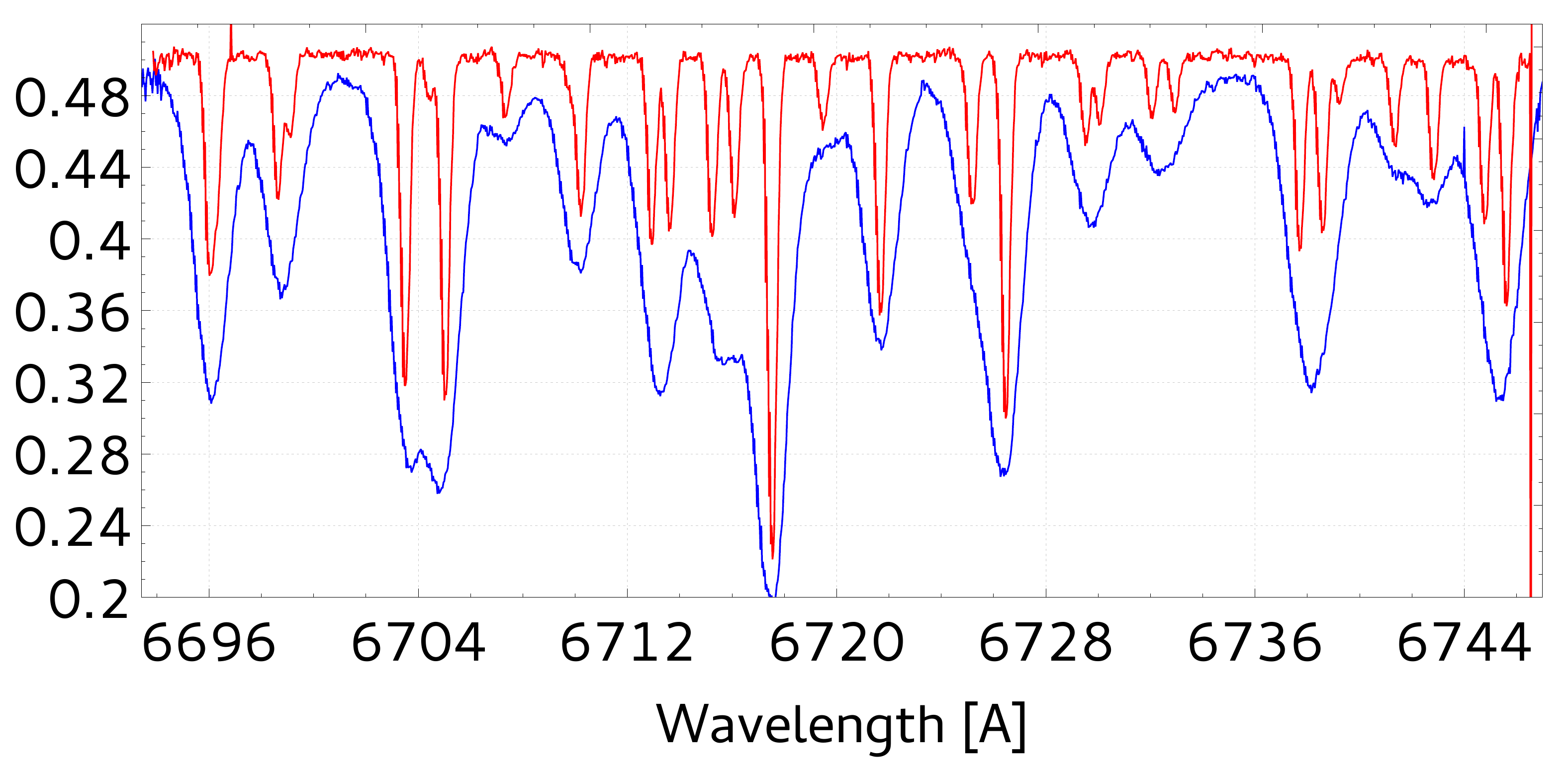}
  \label{Fig11:sfig1}
 \end{subfigure}
\begin{subfigure}{.5\textwidth}
\centering
\caption{Differences}
 \includegraphics[width=85mm]{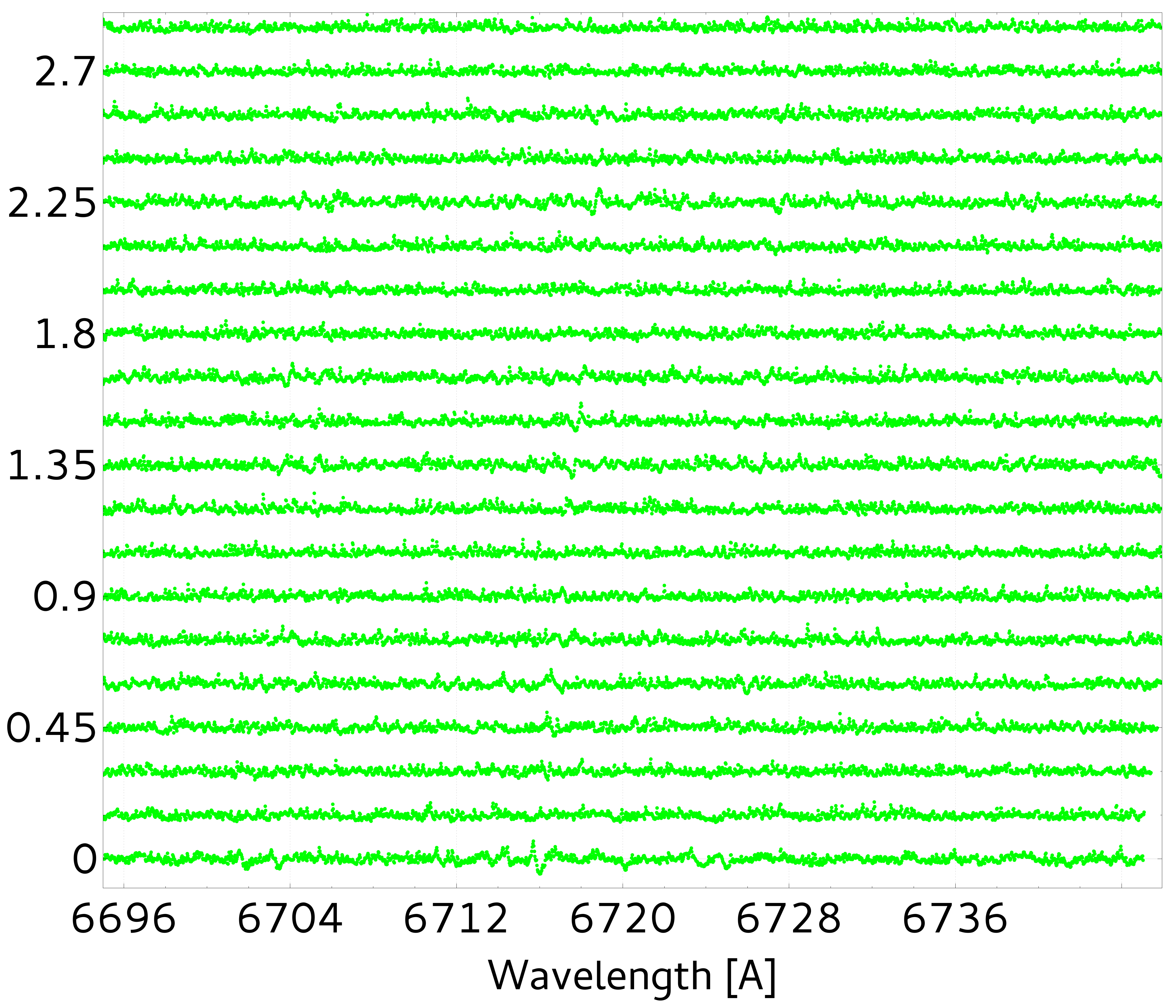}
  \label{Fig11:sfig2}
\end{subfigure} 
\end{center}
 \caption{(a) Disentangling result of our artifical binary with S/N of 100. (b) Differences between result and all spectra.}
  \label{Fig11}
  \end{figure}

We use time-series of artificial spectra with different noise levels (S/N of 10, 20, 40, 60, 80 and 100) to demonstrate the optimisation and error estimation.
One time-series with applied S/N of 100 is shown in Fig. \ref{Fig10}.
The $v \sin(i)$ for the sharp-lined and broad-lined component is 5 km/s and 20 km/s, respectively. The corresponding results from disentangling with best parameters
and the differences between these results and the artificial spectra are shown in Figs. \ref{Fig11:sfig1} and \ref{Fig11:sfig2}, respectively. As mentioned
in Sect. \ref{DSM}, poor definition at the edges can influence the residuum. The plot of the differences in Fig. \ref{Fig11:sfig2} shows this for the lower three spectra.
Hence, the differences are not computed for the whole wavelength range at the red end.

  Table \ref{TabSB2} summarises the data before and after optimisation. The first row lists the S/N values of the result for the sharp-lined component
  spectra measured in two wavelength regions. $\sigma^{opt}_A$ and $\sigma^{opt}_B$ are the summed least squares between calculated RVs and the optimised values for the
  primary and secondary, respectively. We also list the summed least squares, $\sigma_A$ and $\sigma_B$, between calculated RVs and those from the cross-correlation with \textit{CroCo}.
  These values were used to start the optimisation.
  We performed 120 iterations of the optimisation on the RV values and fitted the orbit afterwards. The errors from the fit are also listed in Table \ref{TabSB2}.
  
  \begin{table*}
  \begin{center}
   \caption{Data for SB2 time-series for all 6 data sets. S/N1 and S/N2 is the S/N measured in the result of the sharp-lined component between [$\lambda$6734.2:6737] and [$\lambda$6700:6702.5], respectively.
   The $\sigma$-values are the summed least squares between the calculated RVs and the optimised RVs and between calculated and those from \texttt{CroCo}. Furthermore
   we list all the errors for the orbital parameters before and after optimisation.}
   \begin{tabular}{lllllll}
    \hline \noalign{\smallskip}
    Data: & SN10 & SN20 & SN40 & SN60 & SN80 & SN100\\
    \noalign{\smallskip}\hline \noalign{\smallskip}
    S/N1 / S/N2 & 27/28 & 58/61 & 106/115 & 173/136 & 231/200 & 252/273\\
    $\sigma_A$ [km/s]       & 1.571 & 0.807 & 0.783& 0.775 & 0.745 & 0.752\\
    $\sigma^{opt}_A$ [km/s] & 1.540 & 0.760 & 0.748& 0.680 & 0.643 & 0.520\\
    $\sigma_B$ [km/s]       & 5.930 & 2.779 & 2.166& 2.146 & 2.206 & 2.025\\
    $\sigma^{opt}_B$ [km/s] & 4.146 & 1.831 & 1.566& 1.514 & 1.351 & 1.525\\
    $\delta(P)$ [days] & $ \pm 8.7 \times 10^{-6}$ & $\pm 3.6 \times 10^{-6}  $ & $\pm 2.8 \times 10^{-6} $ & $\pm 3.1 \times 10^{-6} $ & $\pm 2.9 \times 10^{-6} $ & $ \pm 2.5 \times 10^{-6} $\\
    $\delta(P^{opt})$ [days] & $\pm 7.6 \times 10^{-6}$ & $\pm 3.8\times 10^{-6}$ & $ \pm 3.0 \times 10^{-6} $ & $\pm 2.8 \times 10^{-6} $ & $\pm 2.6 \times 10^{-6} $ & $\pm 2.3 \times 10^{-6} $\\
    $\delta(e)$ & $\pm 0.0077$ & $\pm 0.0035$ & $\pm 0.0027 $ & $\pm 0.0031$ & $\pm 0.0029$ & $\pm 0.0025$\\
    $\delta(e^{opt})$ & $\pm 0.0076 $ & $\pm 0.0040 $ & $\pm 0.0030 $ & $\pm 0.0028 $ & $\pm 0.0026 $ & $\pm 0.0023 $\\
    $\delta(K_A)$ [km/s] & $\pm 0.56$ & $\pm 0.26 $ & $\pm 0.21$& $\pm 0.24$ & $\pm 0.22$ & $\pm 0.19$\\
    $\delta(K^{opt}_A)$ [km/s] & $\pm 0.58$ & $\pm 0.31 $ & $\pm 0.25 $& $\pm 0.23 $ & $\pm 0.21 $ & $\pm 0.17 $\\
    $\delta(K_B)$ [km/s] & $\pm 1.9 $ & $\pm 0.78 $ & $\pm 0.52$& $\pm 0.59$ & $\pm 0.58$ & $\pm 0.46$\\
    $\delta(K^{opt}_B)$ [km/s] & $\pm 1.4 $ & $\pm 0.68 $ & $\pm 0.47 $& $\pm 0.46 $ & $\pm 0.41 $ & $\pm 0.45 $\\
    $\delta(\gamma)$ [km/s] & $\pm 0.38$ & $\pm 0.16$ & $\pm 0.13$& $\pm 0.14$ & $\pm 0.14 $ & $\pm 0.11$\\
    $\delta(\gamma^{opt})$ [km/s] & $\pm 0.35 $ & $\pm 0.18 $ & $\pm 0.14 $& $\pm 0.13 $ & $\pm 0.12 $ & $\pm 0.11 $\\
    $\delta(T_0)$ & $\pm 0.00088 $ & $\pm 0.00037$ & $\pm 0.00029$& $\pm 0.00032$ & $\pm 0.00030$ & $\pm 0.00026$\\
    $\delta(T^{opt}_0)$ & $\pm 0.00079 $ & $\pm 0.00040 $ & $\pm 0.00032 $ & $ \pm 0.00029 $& $\pm 0.00027 $ &$\pm 0.00024 $\\
    $\delta(\omega_A)$ & $\pm 0.95$ &$\pm 0.44 $ &$\pm 0.35$& $\pm 0.40$ &$\pm 0.36$ & $\pm 0.32$\\
    $\delta(\omega^{opt}_A)$ & $\pm 0.95 $ &$\pm 0.51 $ &$\pm 0.39 $ &$\pm 0.36 $ & $\pm 0.34 $ &$\pm 0.29 $\\
    $\delta(M_A \sin^3i)$ [$M_{\sun}$] & $\pm 0.00081$ &$\pm \pm 0.00037$ &$\pm 0.00025$& $\pm 0.00029$ &$\pm 0.00028$ &$\pm 0.00022$\\
    $\delta(M^{opt}_A \sin^3i)$ [$M_{\sun}$]& $\pm 0.00064$ &$\pm 0.00034$ &$\pm 0.00024$& $\pm 0.00023$ &$\pm 0.00021$ &$\pm 0.00022$\\
    $\delta(M_B \sin^3i)$ [$M_{\sun}$]& $\pm 0.00054$ &$\pm 0.00024$ &$\pm 0.00018$& $\pm 0.00020$ &$\pm 0.00019$ &$\pm 0.00016$\\
    $\delta(M^{opt}_B \sin^3i)$ [$M_{\sun}$]& $\pm 0.00048$ &$\pm 0.00025$ &$\pm 0.00019$& $\pm 0.00018$ &$\pm 0.00016$ &$\pm 0.00015$\\
    \noalign{\smallskip} \hline
   \end{tabular}
  \label{TabSB2}
  \end{center}
  \end{table*}

  In all cases, the resulting spectra show a significantly higher S/N compared to the input data by at least a factor of two.
  Additionally, the resulting RVs from cross-correlation and from optimisation of disentangling improve with increasing S/N.
  From the errors for the masses of the components, which we are most likely interested in, we can see a significant reduction in the errors.
  These errors could be further reduced by continuing with the optimisation. We note that we used ideal input data, equally normalised,
  identical quality and equidistantly spread over the orbital phase. This itself yields good results for the cross-correlation method. However,
  disentangling further improved the results.
  
  \begin{figure}[h!]
 \begin{center}
 \includegraphics[width=85mm]{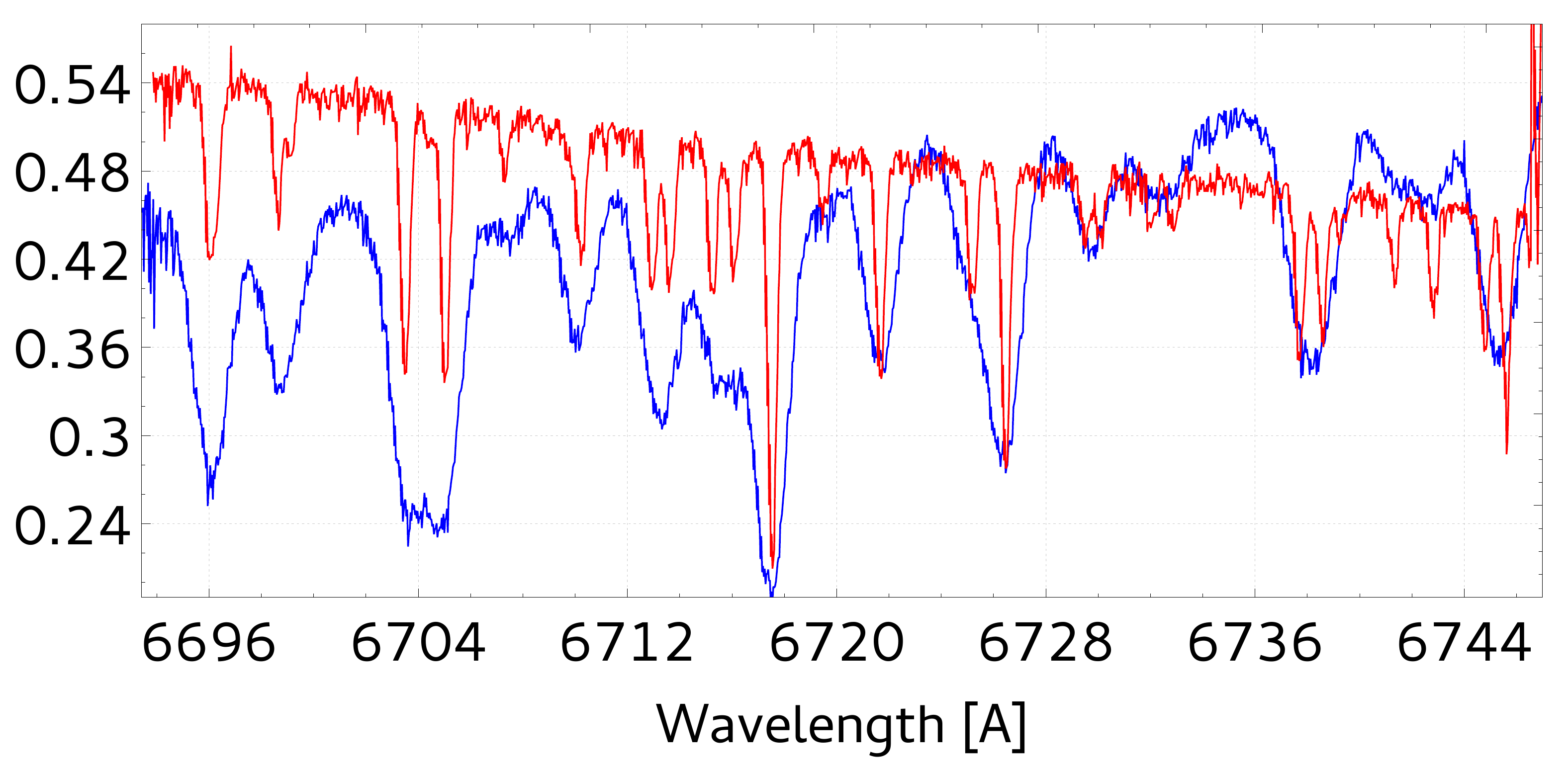}
\end{center}
 \caption{Result from disentangling of the SN100 set with two spectra replaced from the SN10 set. The continuum is less well determined and the noise
 in the result is increased.}
  \label{FigNoise}
  \end{figure}
  
    \begin{figure}[h!]
 \begin{center}
 \includegraphics[width=85mm]{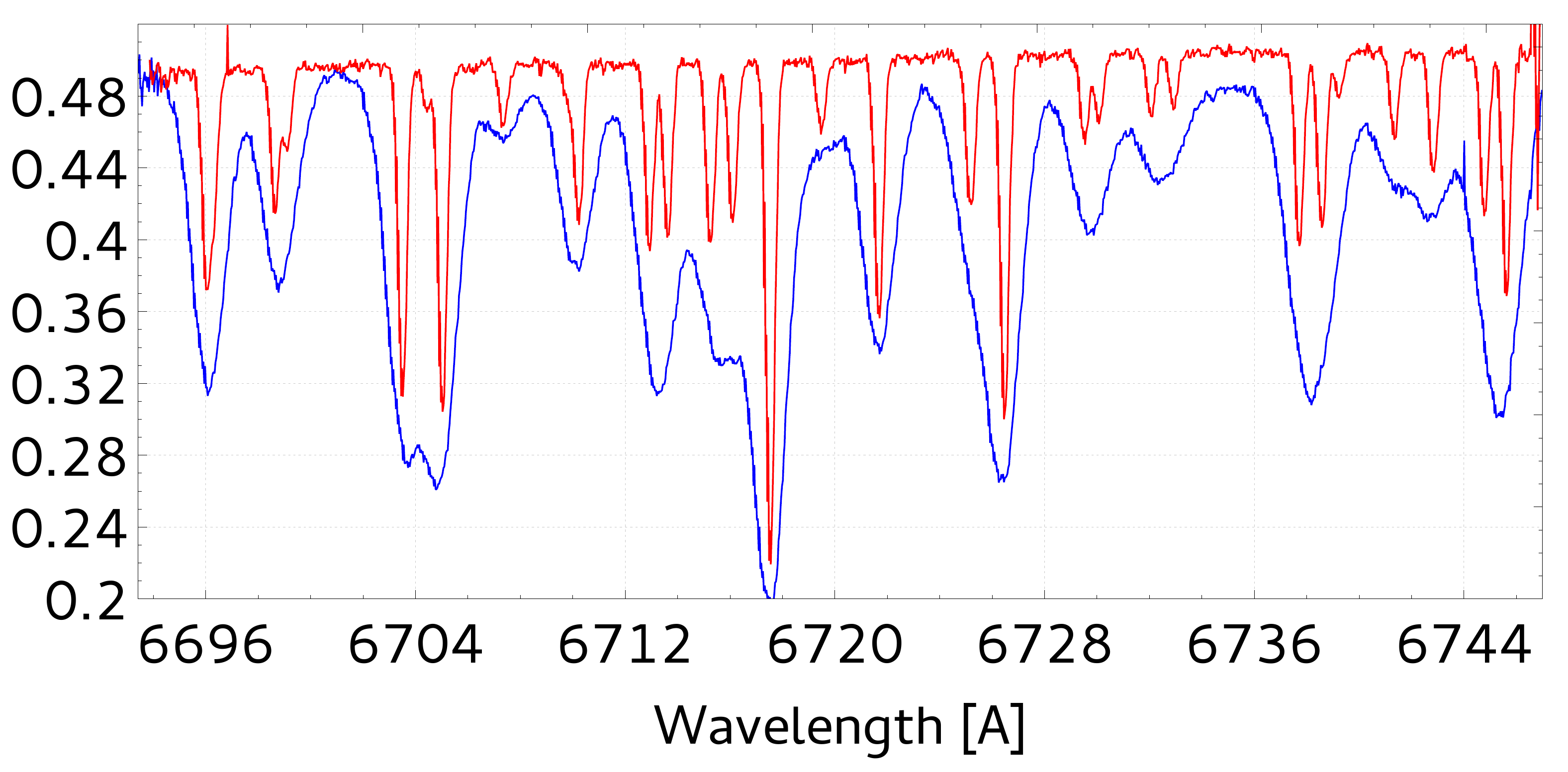}
\end{center}
 \caption{Result from the disentangling of SN100 set with three spectra on different continuum levels. Compared to the results in Fig.~\ref{Fig11:sfig1},
 there is only a slight change in the continuum level of the results.}
  \label{FigNorm}
  \end{figure}
  
  The way that unequal noise-levels in a time-series influence the result is shown in Fig.~\ref{FigNoise}. We have replaced two spectra from the SN100 set by those from the SN10 set. The continuum is now less well determined and the noise in the resulting spectra is increased (compared to Fig.~\ref{Fig11:sfig1}).
  Only a few spectra with significant higher noise will make line-profile analysis difficult.
  Additionally, in Fig.~\ref{FigNorm} we show the result for unequal normalisation levels. Therefore, we have set the continuum of spectrum 2 to 0.98, spectrum 14
  to 1.05 and spectrum 19 to 0.9. There is only a slight change in the continuum level in the results.
  
  \begin{figure}[h!]
 \begin{center}
  \begin{subfigure}{.5\textwidth}
  \centering
  \caption{For component 1}
  \includegraphics[width=85mm]{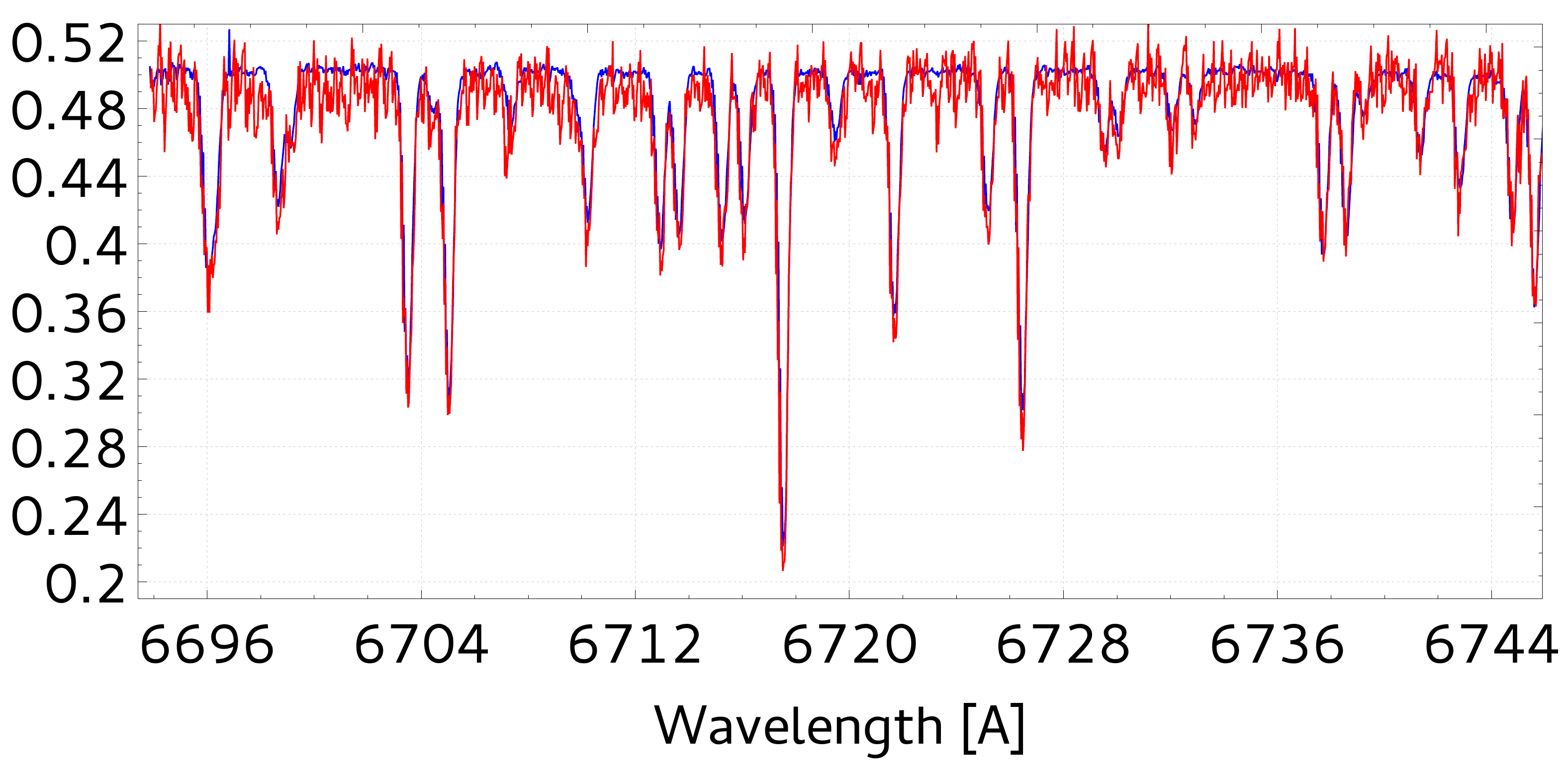}
  \label{FigMin:sfig1}
 \end{subfigure}
\begin{subfigure}{.5\textwidth}
\centering
\caption{For component 2}
 \includegraphics[width=85mm]{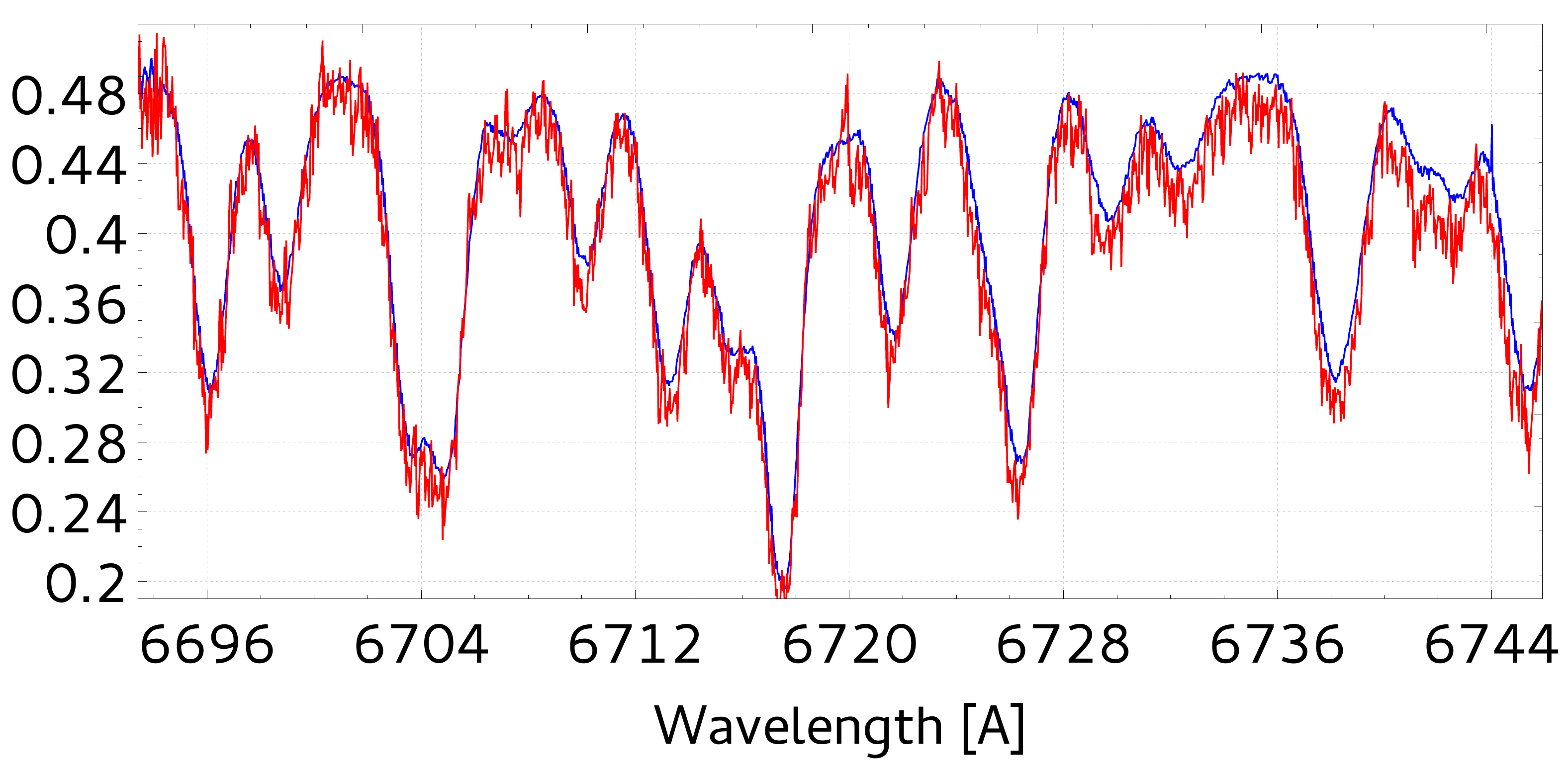}
  \label{FigMin:sfig2}
\end{subfigure} 
\end{center}
 \caption{Overplot of results for SN100 minimum case (three spectra) and all (20 spectra). (a) for the narrow-lined component and (b) for the broad-lined component.}
  \label{FigMin}
  \end{figure}
  
  Furthermore, we show a comparison of the disentangling result between the minimum case, where we used three spectra, and the result from all
  20 spectra in Fig.~\ref{FigMin}. We see that the contributions of the components are correctly separated. However, the noise level in the
  result is high and makes detailed line-profile analysis difficult.
  
\section{Application to Capella}
\label{sec:application}

This system is somehow unique owing to its very precisely determined masses \citep{2011A&A...531A..89W} and evolutionary state \citep{0004-637X-807-1-26}.
From our STELLA \citep[see][]{doi:10.1117/12.790687} telescopes on Tenerife, a lot of spectroscopic data is available for Capella. We selected some for our disentangling study, which are listed in Table~\ref{tab2}.

\begin{table}
\caption{Spectra used for the studies in Sect. \ref{sec:application}. The x in the last three columns indicate which spectra are used in the respective subsections. The S/N of these spectra is $\gtrsim$ 100.}
\begin{tabular}{lllllll}
\hline  \noalign{\smallskip}
No. & HJD 2456+ & RV A & RV B & Sect. & & \\
 ~ & & [km/s] & [km/s] & \ref{minimum} & \ref{quadratur} & \ref{real} \\
 \noalign{\smallskip}\hline  \noalign{\smallskip}
0 & 347.591 & 34.97 & 25.01 & ~ & ~ & x\\
1 & 364.534 & 54.33 & 4.82 & x & x & x\\
2 & 366.344 & 55.20 & 3.92 & ~ & x & x\\
3 & 366.520 & 55.22 & 3.86 & ~ & x & ~\\
4 & 369.340 & 55.86 & 3.15 & ~ & x & ~\\
5 & 370.387 & 55.91 & 3.08 & ~ & x & ~\\
6 & 371.341 & 55.86 & 3.12 & ~ & x & ~\\
7 & 372.341 & 55.72 & 3.25 & ~ & x & ~\\
8 & 373.246 & 55.48 & 3.48 & ~ & x & ~\\
9 & 377.343 & 53.62 & 5.34 & x & x& x\\
10 & 378.526 & 52.78 & 6.17 & ~ & x & x\\
11 & 379.523 & 52.01 & 6.97 & ~ & x & ~\\
12 & 381.344 & 50.38 & 8.63 & ~ & x & x\\
13 & 381.524 & 50.22 & 8.81 & ~ & x & ~\\
14 & 382.529 & 49.18 & 9.86 & ~ & x & ~\\
15 & 383.515 & 48.11 & 10.95 & & x & x\\
16 & 398.420 & 26.72 & 32.99 & x & x & x\\
17 & 399.412 & 25.19 & 34.57 & ~ & ~ & x\\
18 & 414.369 & 6.95 & 53.57 & ~ & x & x\\
19 & 415.368 & 6.29 & 54.27 & ~ & x & ~\\
20 & 416.369 & 5.69 & 54.91 & ~ & x & ~\\
21 & 417.386 & 5.17 & 55.46 & ~ & x & ~\\
22 & 418.370 & 4.76 & 55.91 & ~ & x & ~\\
23 & 420.364 & 4.19 & 56.53 & ~ & x & ~\\
24 & 515.669 & 9.26 & 51.15 & ~ & x & x\\
25 & 518.661 & 6.78 & 53.75 & ~ & x & x\\
26 & 528.663 & 4.23 & 56.56 & x & x & x\\
27 & 531.621 & 5.26 & 55.54 & x & x & x\\
28 & 538.635 & 10.73 & 49.98 & ~ & ~ & x\\
29 & 544.605 & 18.14 & 42.39 & ~ & ~ & x\\
30 & 549.752 & 25.78 & 34.51 & ~ & ~ & x\\
31 & 570.578 & 53.05 & 6.17 & ~ & ~ & x\\
\noalign{\smallskip} \hline
\end{tabular}
\label{tab2}
\end{table}

\begin{figure}[h!]
 \begin{center}
 \begin{subfigure}{.5\textwidth}
  \centering
  \caption{Data for the ideal minimum case}
  \includegraphics[width=85mm]{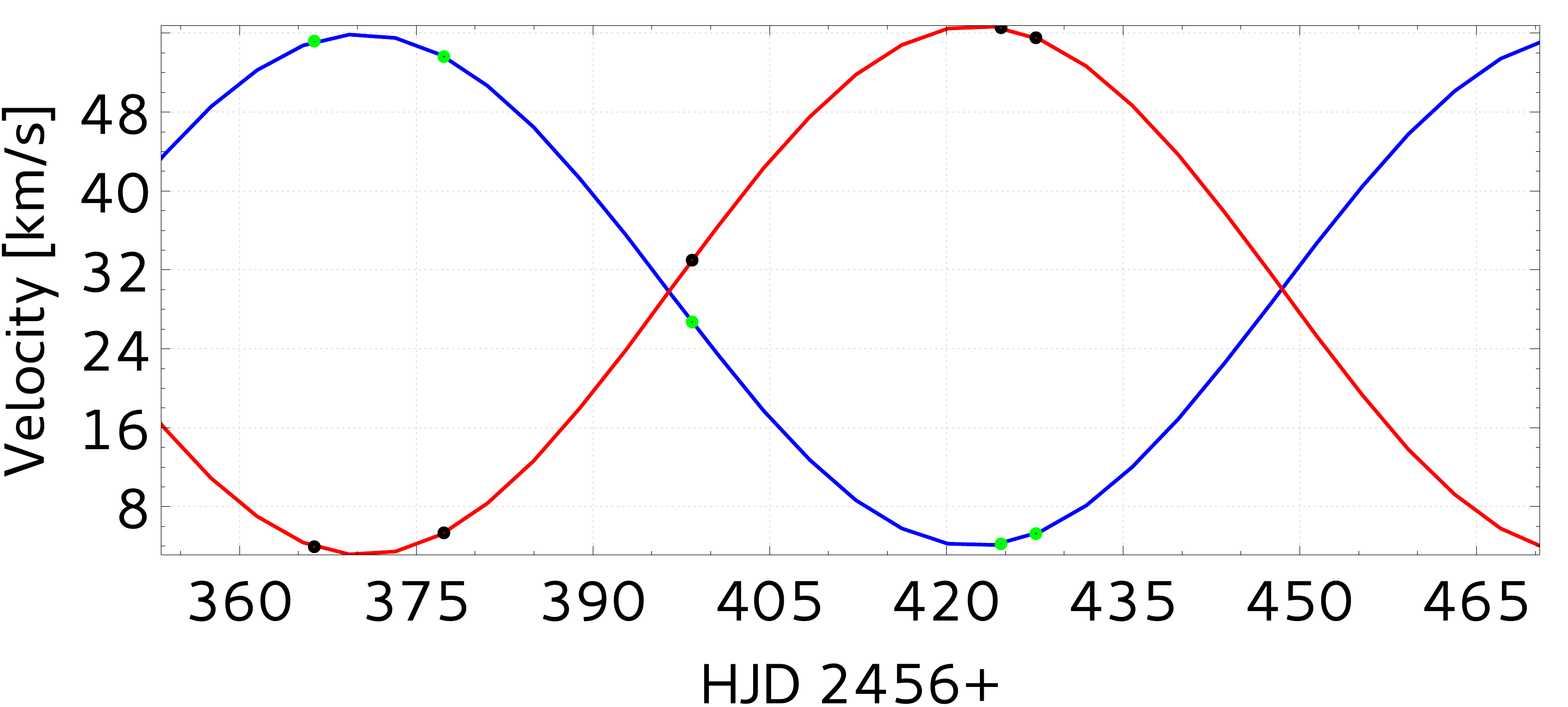} 
  \label{Fig12:sfig1}
 \end{subfigure}
 \begin{subfigure}{.5\textwidth}
  \centering
  \caption{Data at quadrature (modulo period to spectrum no. 1)}
  \includegraphics[width=85mm]{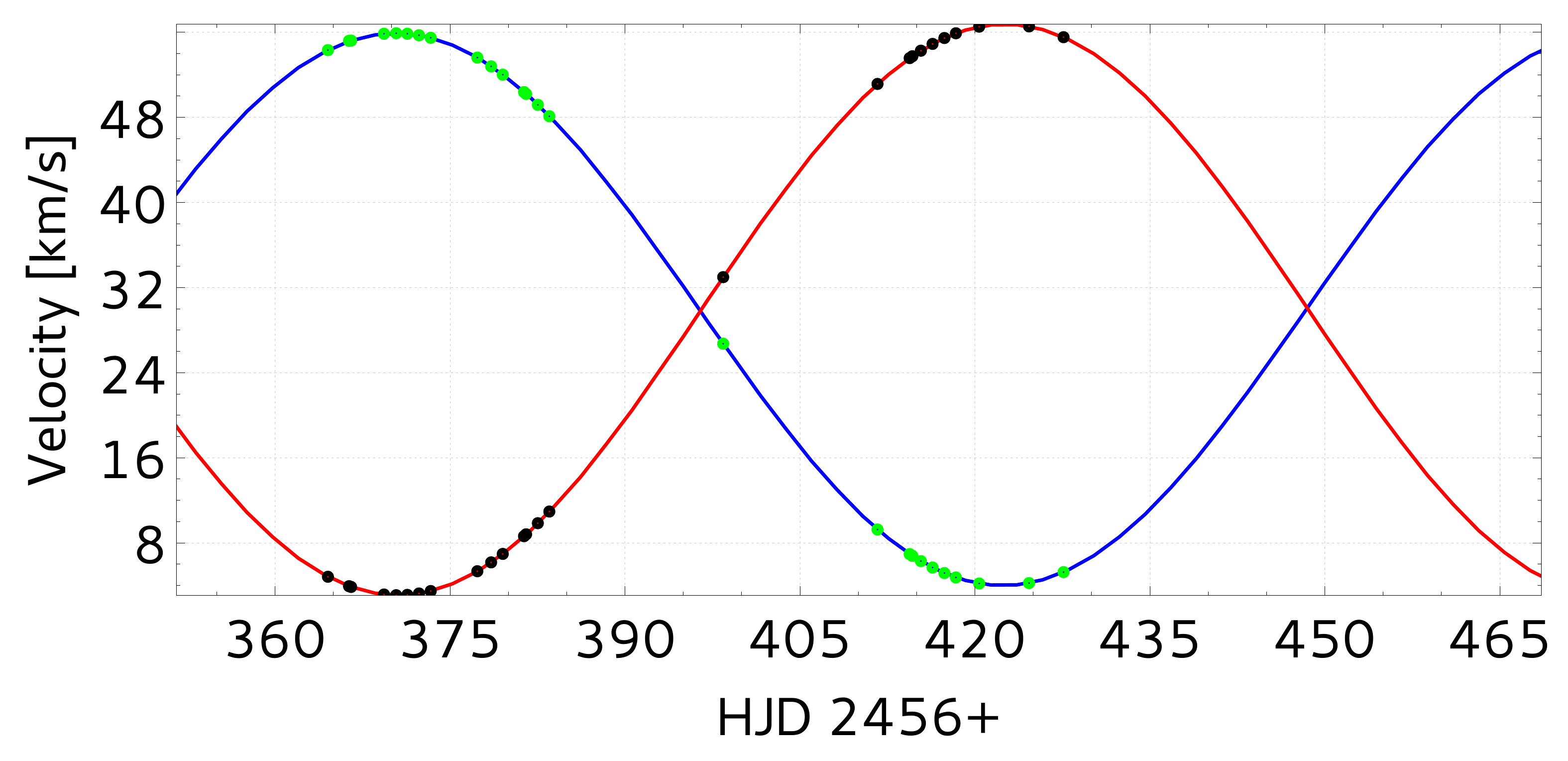} 
  \label{Fig12:sfig2}
 \end{subfigure}
 \begin{subfigure}{.5\textwidth}
  \centering
  \caption{Data for homogeneous coverage (modulo period to spectrum no. 0)}
  \includegraphics[width=85mm]{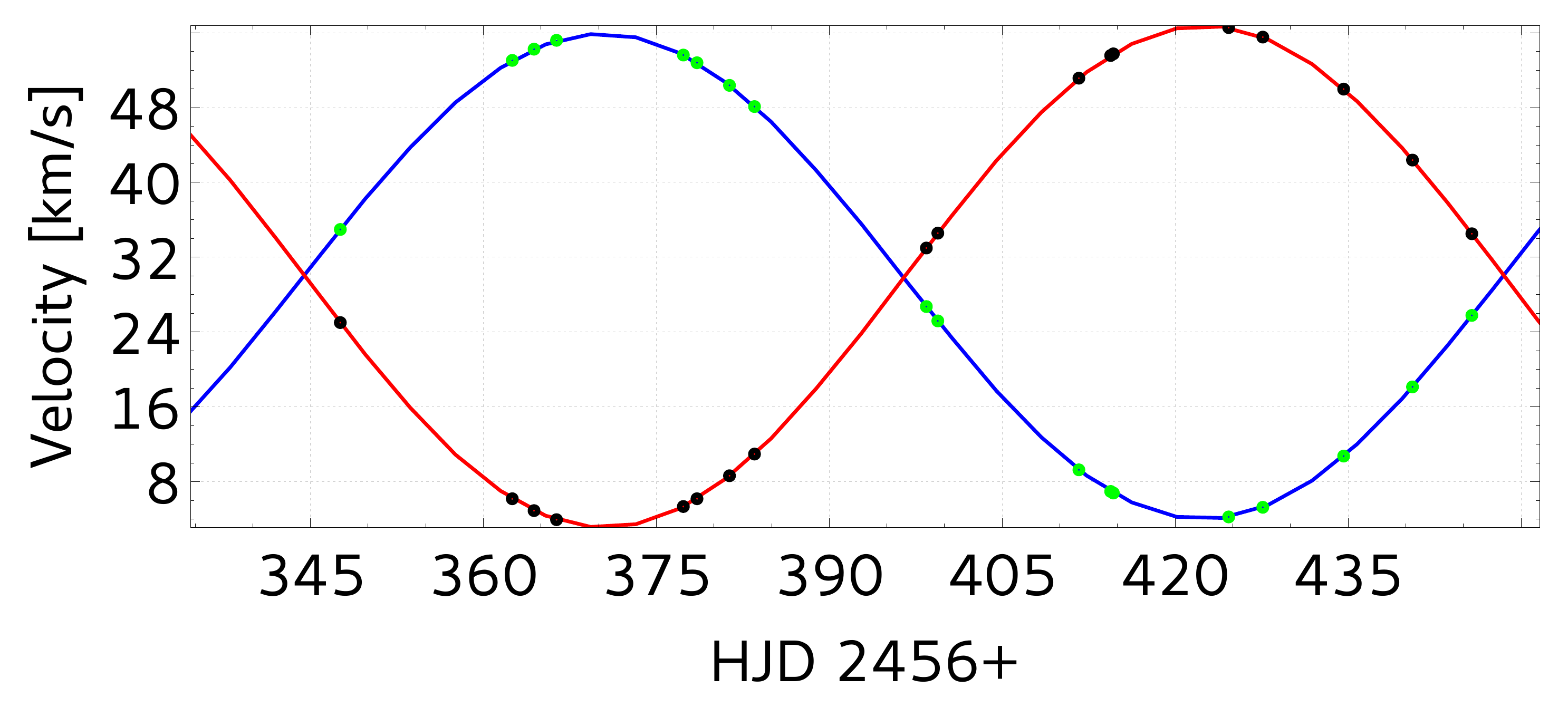} 
  \label{Fig12:sfig3}
 \end{subfigure}
\end{center}
\caption{Radial velocities from Table~\ref{tab2} for the three special cases studied. The blue (red) curve show the primary's (secondary's) orbital motion (computed from orbital parameters from \citet{0004-637X-807-1-26}), green (black) dots correspond to the primary (secondary) radial velocity of each spectrum.}
  \label{Fig12}
  \end{figure}

\subsection{The ideal minimum case}
\label{minimum}

\begin{figure}[h!]
 \begin{center}
 \includegraphics[width=85mm]{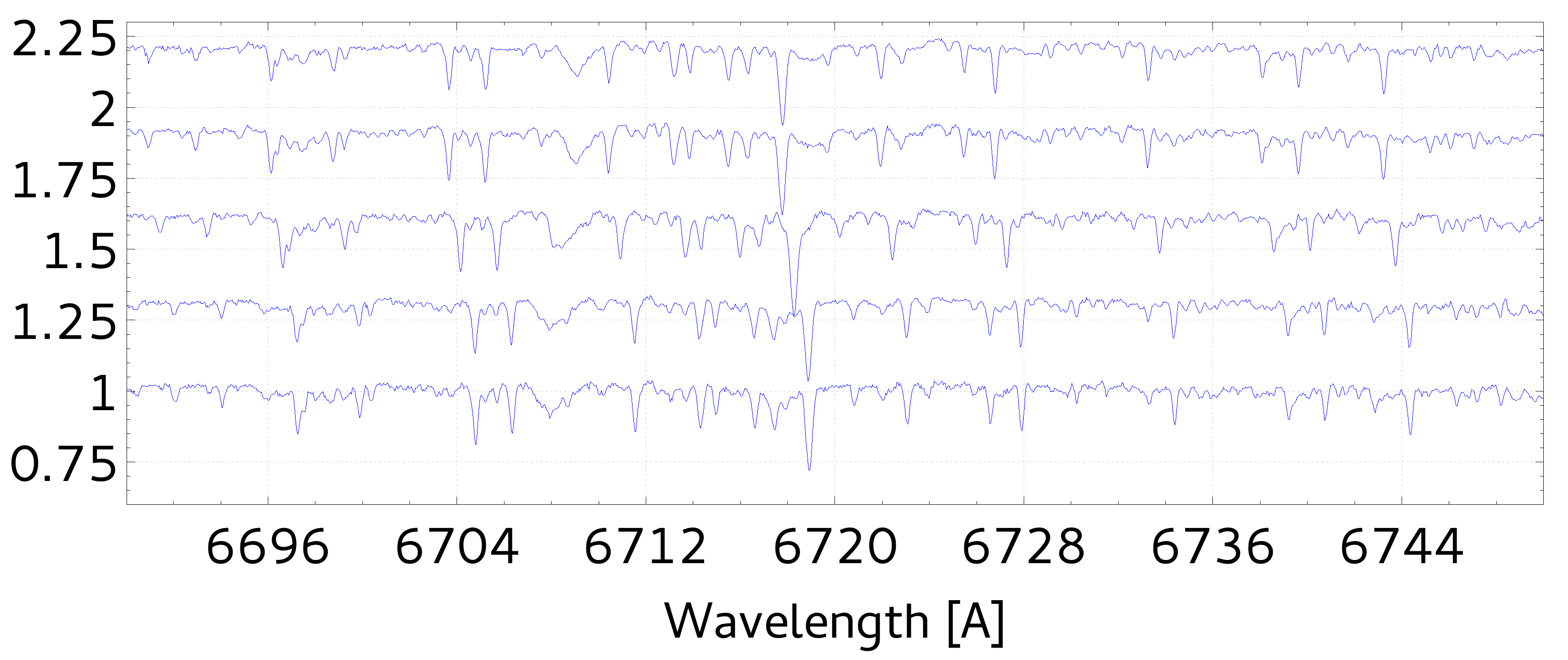}
\end{center}
 \caption{The five spectra from Capella around the \ion{Li}{i}\,6708{\AA} line used for our investigations for the minimum ideal case. Spectrum 1 at the bottom, 5 at the top.}
 \label{Fig13}
  \end{figure}

To show the reliability of the code, we decided to use only a small data set.
As described, as long as the SVD is applied to overdetermined and rank deficient systems it yields the solution of smallest residuum. The system of a binary without telluric lines has therefore two unknowns and, to fulfil the criterion of overdetermination, we need three independent spectra. The time-series of spectra is shown in Fig.~\ref{Fig13}  as well as the radial velocity curve with the measurements in Fig.~\ref{Fig12:sfig1}. We used different combinations of the spectra 2, 9, 16, 26, and 27
as listed in Table~\ref{tab2}. The two additional spectra are used to show how they can help to reduce the noise in the result.

\begin{figure}[h!]
 \begin{center}
  \begin{subfigure}{.5\textwidth}
  \centering
  \caption{}
  \includegraphics[width=85mm]{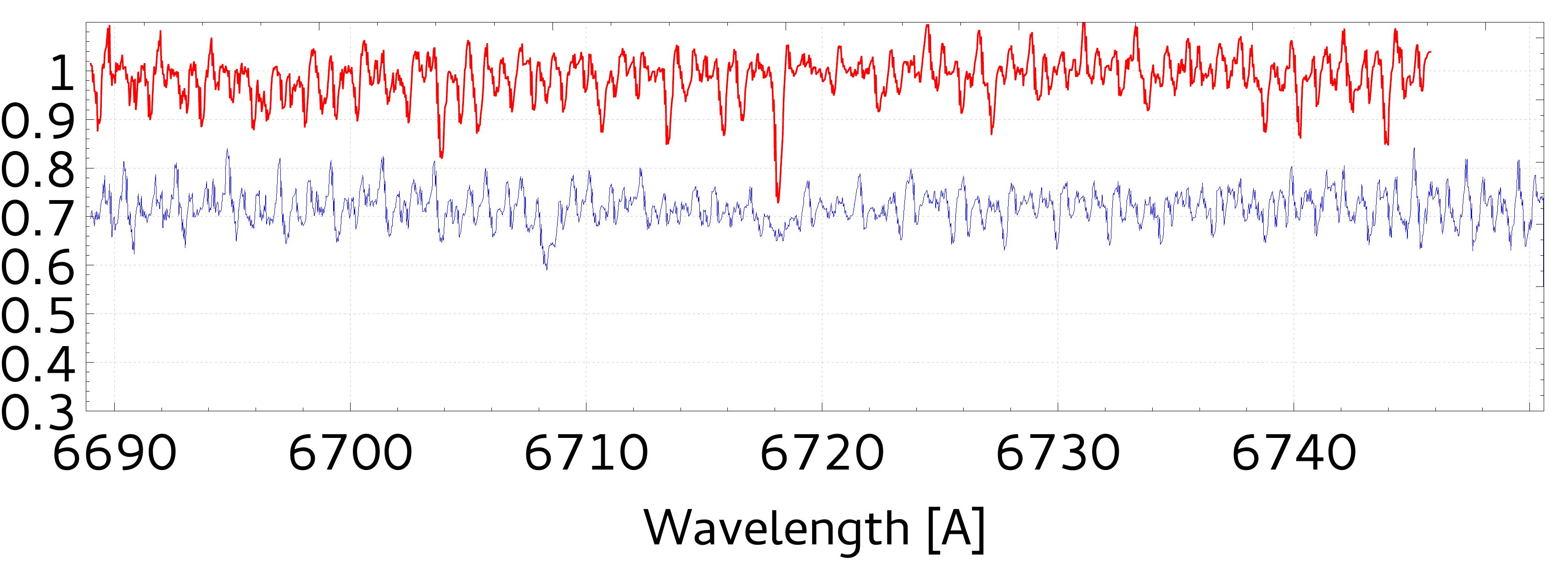}
  \label{Fig14:sfig1}
 \end{subfigure}
\begin{subfigure}{.5\textwidth}
\centering
\caption{}
 \includegraphics[width=85mm]{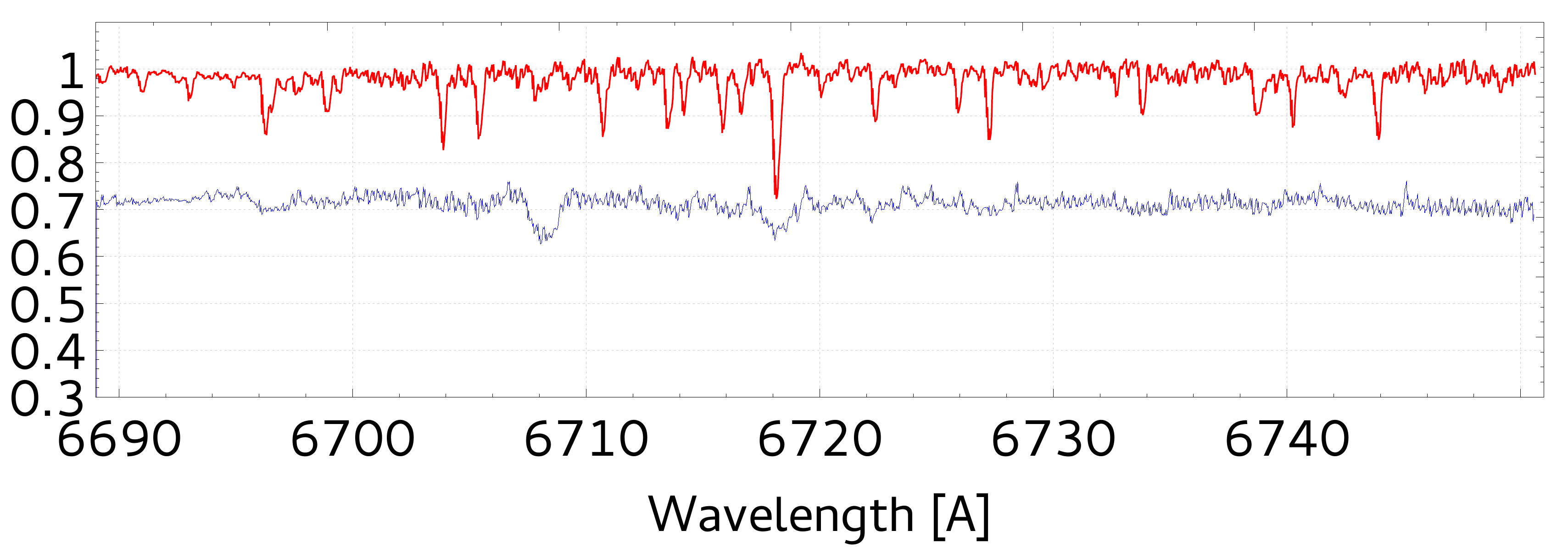}
  \label{Fig14:sfig2}
\end{subfigure}
\begin{subfigure}{.5\textwidth}
  \centering
  \caption{}
  \includegraphics[width=85mm]{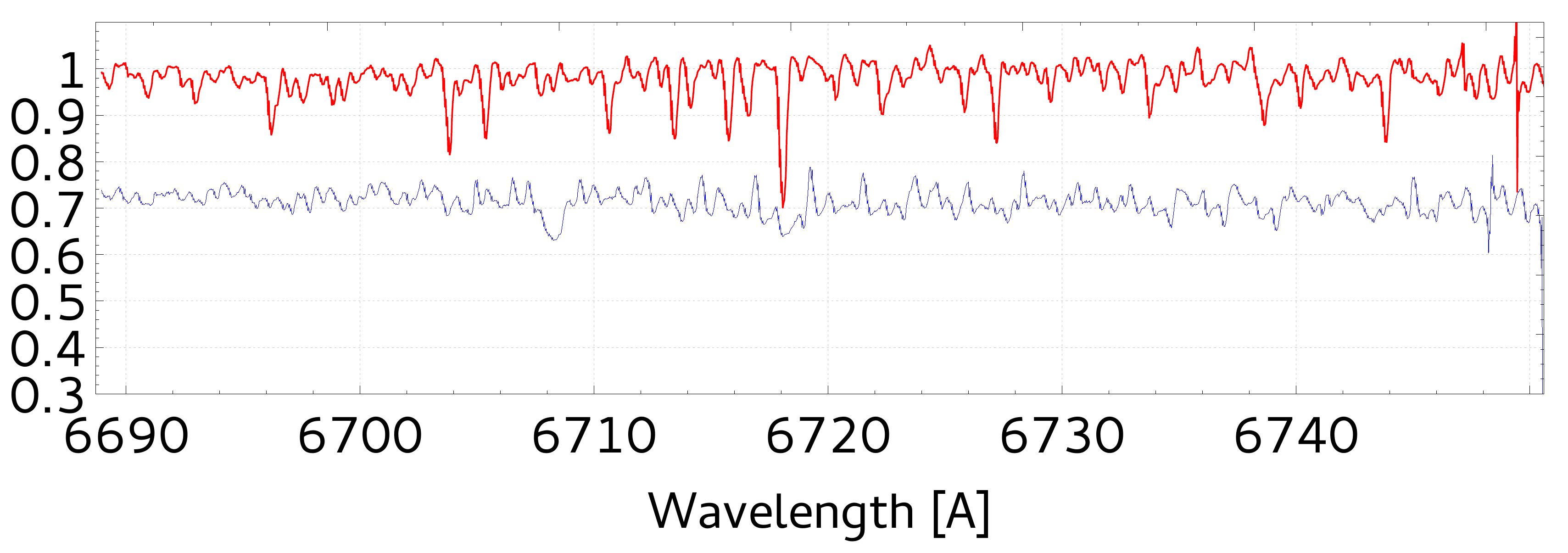}
  \label{Fig14:sfig3}
 \end{subfigure}
\begin{subfigure}{.5\textwidth}
\centering
\caption{}
 \includegraphics[width=85mm]{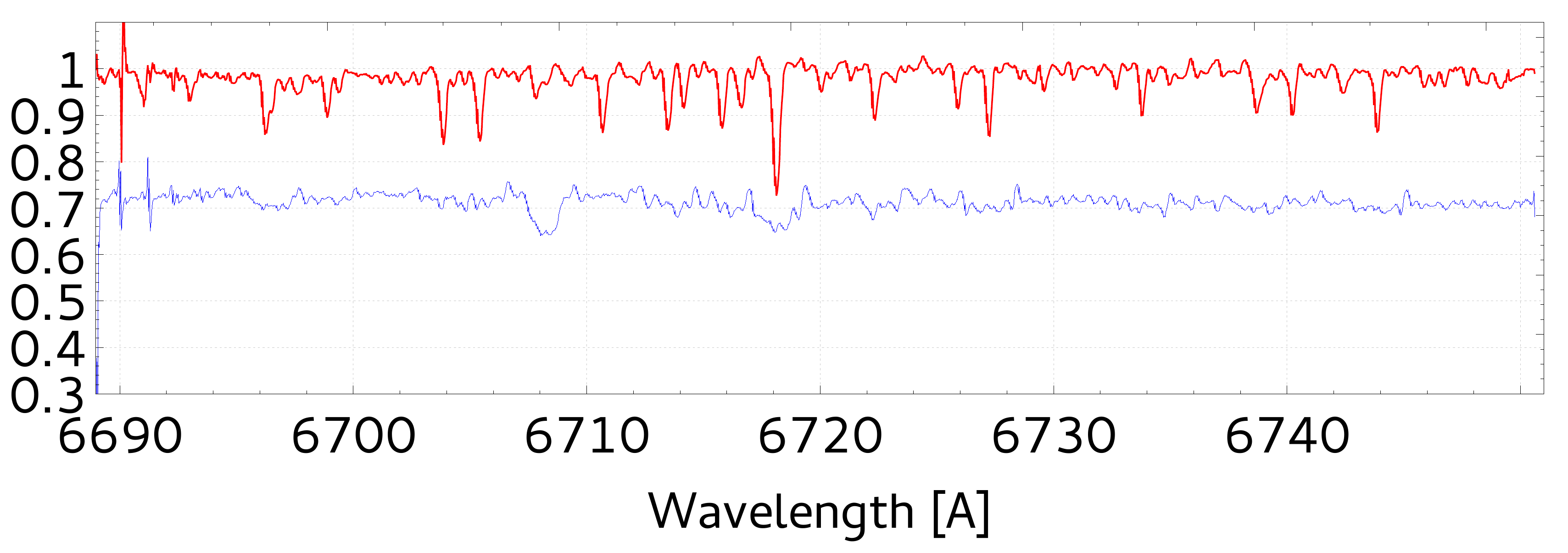}
  \label{Fig14:sfig4}
\end{subfigure} 
\end{center}
 \caption{Results from disentangling around the  \ion{Li}{i}\,6708{\AA} line of Capella from the observations indicated in Table~\ref{tab2}.
 (a): Result for the determined situation with one spectrum per two consecutive quadratures;
 (b): As above but with an additional spectrum at conjunction, which is the minimal necessary set of required data;
 (c): Two spectra per two consecutive quadratures. This set yields only a determined system, which can distinguish better between noise and the signal of the components; 
 (d): All five spectra yield a minimum overdetermined system with good noise reduction.}
  \label{Fig14}
  \end{figure}

With these data, we study the minimum case for disentangling. The determined case where we used two spectra from following quadratures (spectra 9 and 26 in Table~\ref{tab2}) is shown in Fig.~\ref{Fig14:sfig1}. Owing to the noise in the data it is not possible for the code to distinguish fully between the two components. Only strong lines in the secondary spectrum (blue) like the \ion{Li}{i} itself and \ion{Ca}{i} $\lambda$6718{\AA} are visible. The minimum case is shown in Fig.~\ref{Fig14:sfig2}, where we included an observation near conjunction (spectrum 16 in
Table~\ref{tab2}). These three spectra yield an overdetermined system and the solution is less affected by noise than is the case for the only determined system.

The next example (Fig.~\ref{Fig14:sfig3}) shows the result when all four spectra from the two consecutive quadratures are used (spectra 1, 2, 26 and 27 in Table \ref{tab2}). If we compare this result with the one above, we can see that the jitter of the solution as a result of noise in the measurements is reduced. However, since the binning of these spectra corresponds to a velocity of 1\,km/s, the two pixel resolution is around 2\,km/s. A glance on the RV values in Table \ref{tab2} shows that the two observations for each pair differ by less than 2\,km/s. This will give the code a better estimate of the noise but not of the components, since we would need three independent conditions to get an overdetermined system.
Figure~\ref{Fig14:sfig4} shows the result when we use all five spectra. Compared to the minimum case (Fig.~\ref{Fig14:sfig2}), the noise is greatly reduced. 

\subsection{Data at quadrature - a non-ideal case}
\label{quadratur}

Observations at maximum separation of the components lead to minimum blending of equal lines of both components and are, therefore, often used to study the properties of the individual components of spectroscopic binaries.
The orbital distribution of the observations we chose for this test (indicated in the sixth column of Table~\ref{tab2}) are shown in Fig. \ref{Fig12:sfig2}. Spectrum 16 is near conjunction and all the others of the 25 spectra are at quadrature. 

We studied six different cases: (1) The first one uses the spectra 1 to 15 from the first quadrature. (2) The second case is with spectra from both quadratures from 1 to 15, and 18 to 27. (3) The third case uses spectra from 1 to 16, i.e., all spectra from one quadrature, plus an additional spectrum near conjunction. (4) For the fourth case we used all 26 spectra. (5) For the fifth case, we use all spectra from the first quadrature and
 spectrum 22 from the following quadrature. (6) As per (5), but with spectrum 16 near conjunction. Since most of the spectra from the same quadrature differ little in their radial
velocities, the overdetermination is weak for cases (1) and (2). As shown for the minimum ideal case, the spectrum near conjunction is advantageous to fulfil the criterion of an overdetermined system.

Spectrum 5 has the highest (lowest) RV for component A (B). The next spectra, which have an RV difference of at least $\approx$~2\,km/s are 9, 12, and 15, i.e., these 15 spectra lead to a weakly overdetermined system.
Two spectra for the determined case, another two spectra for overdetermination, and the rest help to reduce the noise in the result (case 1). The number of independent spectra, $N_{IS}$, for each case is listed in Table~\ref{tab3}. 
  
\begin{figure}[h!]
 \begin{center}
 \includegraphics[width=85mm]{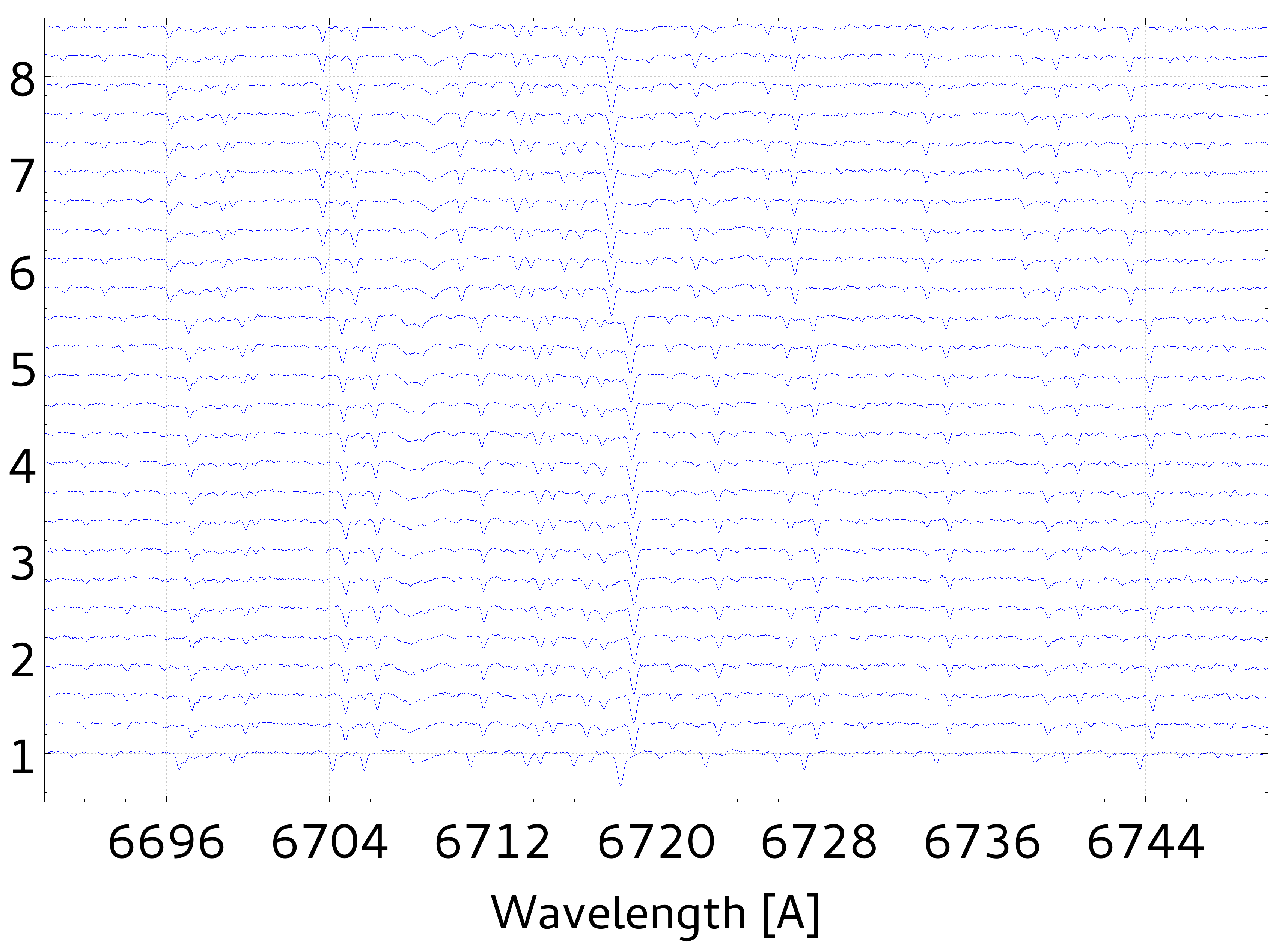}
\end{center}
 \caption{All the spectra from the non-ideal case at quadrature. The spectrum at the bottom is spectrum 16 near conjunction and all others are in the same order as in Table~\ref{tab2}, from bottom to top.}
  \label{Fig16}
  \end{figure}
  
\begin{table}
\caption{Data for the non-ideal case at quadrature. We list the time for a single SVD on 4 cores, the residuum $r$ before/after optimisation, the size of the transformation matrix $\underline{M}$, the number of independent spectra $N_{IS}$ and our quality factor $Q_d$.}
\begin{tabular}{llllll}
\hline  \noalign{\smallskip}
Case & Time & $r$ & $\underline{M}$ & $N_{IS}$ & $Q_d$\\
 ~ & (min) & $10^{-4}$ & $u\times v$ & ~ & ~\\
 \noalign{\smallskip}\hline  \noalign{\smallskip}
 1 & 7.7 & 0.94/0.94 & 5752$\times$42120 & 4 & -0.021\\
 2 & 12.4 & 1.24/0.95 & 5488$\times$70225 & 8 & -0.078\\
 3 & 8.4 & 0.94/0.94 & 5798$\times$44928 & 5 & -0.018\\
 4 & 13.2 & 1.48/1.45 &5848$\times$73034& 9 & -0.096\\
 5 & 8.6 & 0.91/0.89 & 5840$\times$44924& 5& -0.018\\
 6 & 9.7 & 0.91/0.91 & 5840$\times$47736 & 6 & -0.013\\
 \noalign{\smallskip} \hline
\end{tabular}
\label{tab3}
\end{table}

\begin{figure}[h!]
 \begin{center}
  \begin{subfigure}{.5\textwidth}
  \centering
  \caption{Case 1}
  \includegraphics[width=85mm, height = 30mm]{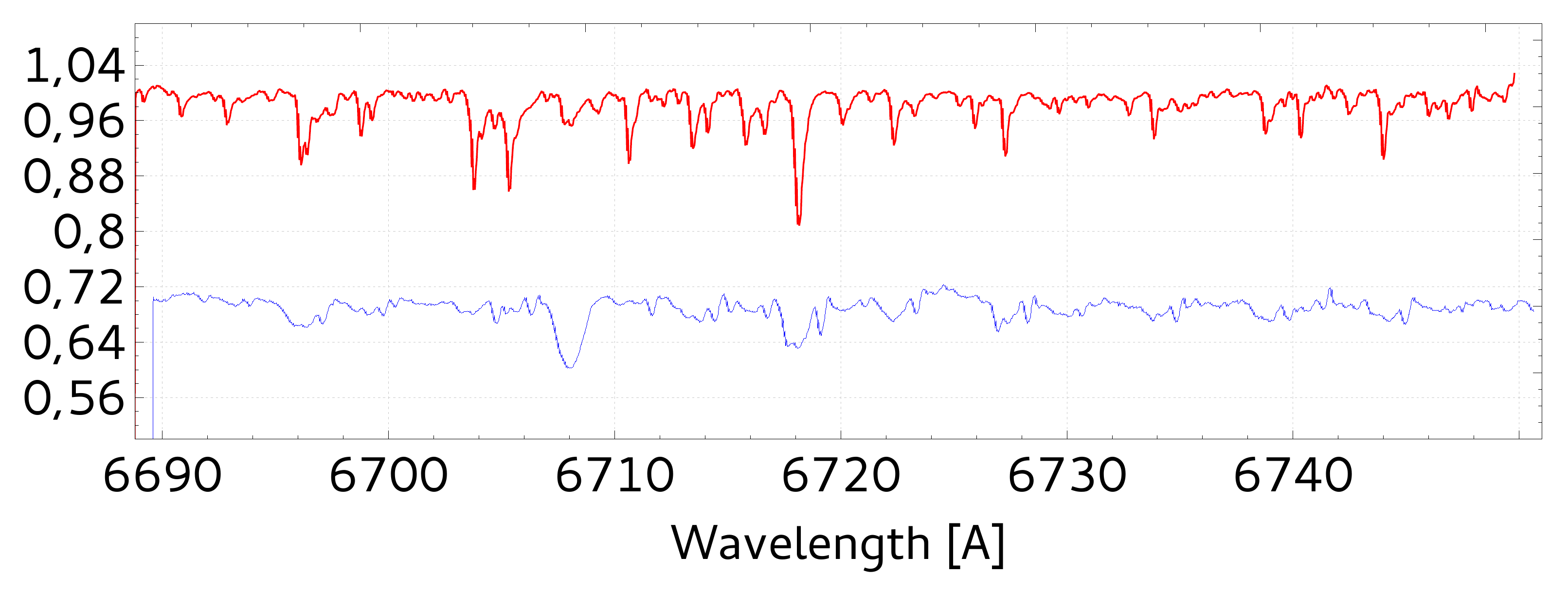}
  \label{Fig17:sfig1}
 \end{subfigure}
 \begin{subfigure}{.5\textwidth}
  \centering
  \caption{Case 2}
  \includegraphics[width=85mm, height = 30mm]{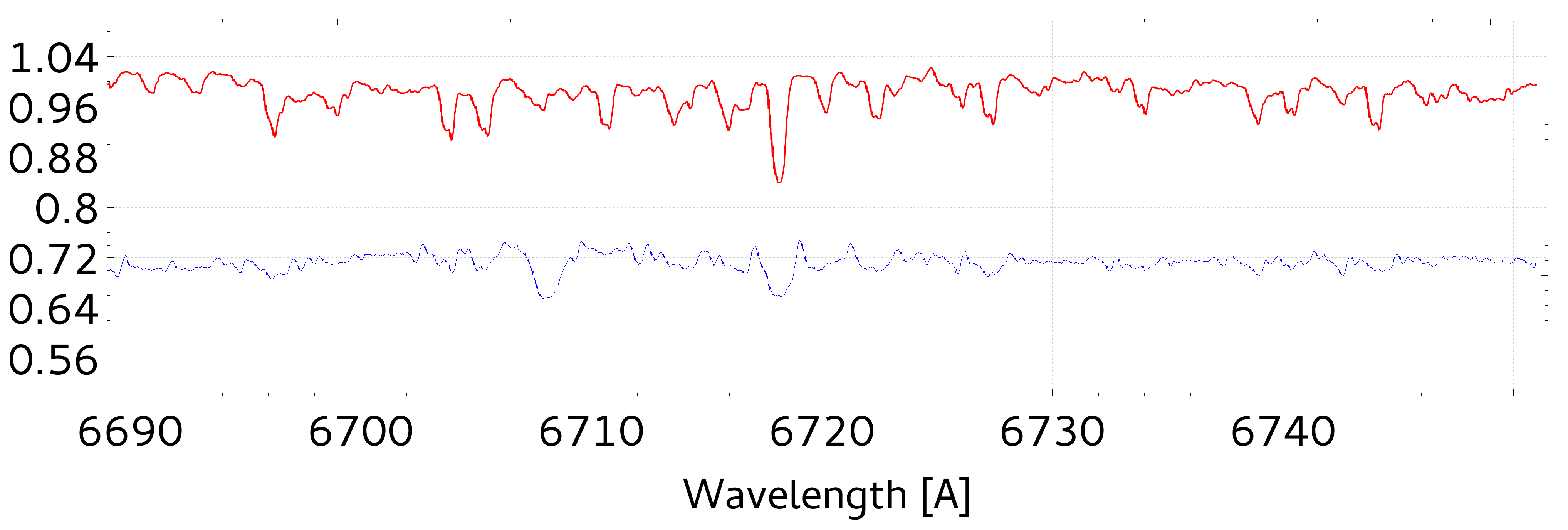}
  \label{Fig17:sfig2}
 \end{subfigure}
  \begin{subfigure}{.5\textwidth}
  \centering
  \caption{Case 3}
  \includegraphics[width=85mm, height = 30mm]{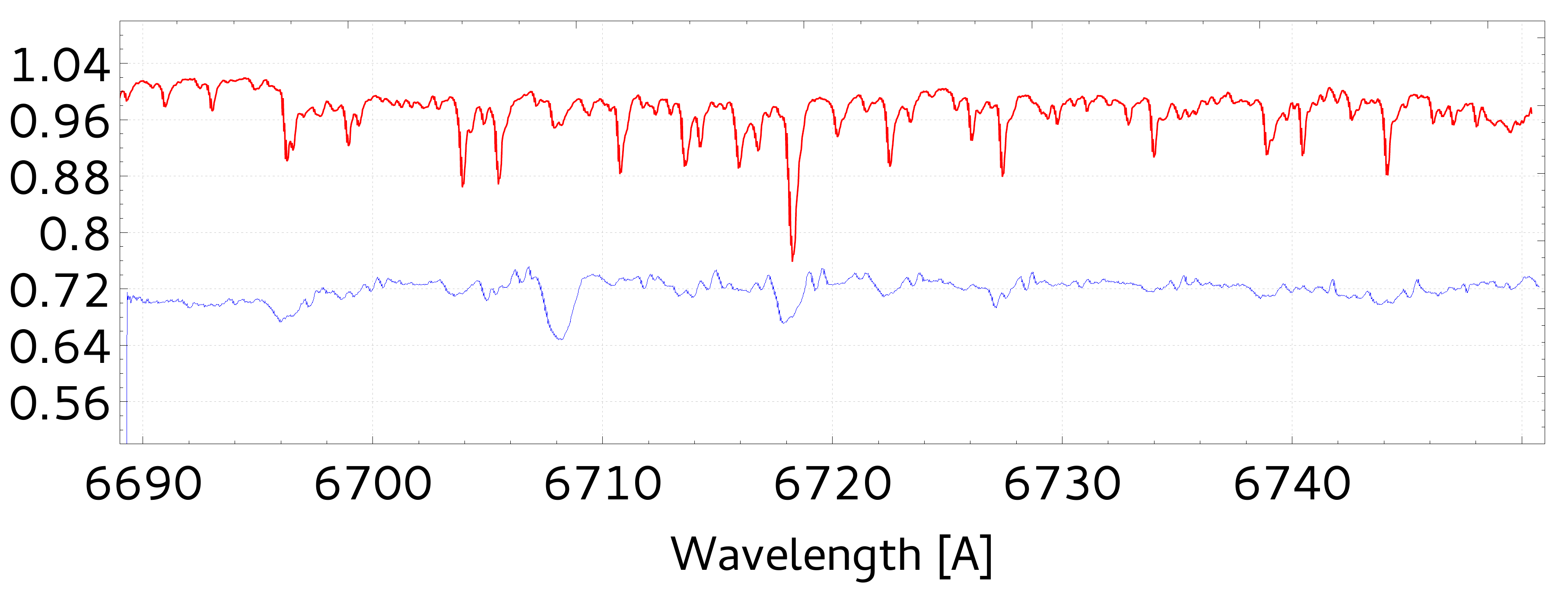}
  \label{Fig17:sfig3}
 \end{subfigure}
 \begin{subfigure}{.5\textwidth}
  \centering
  \caption{Case 4}
  \includegraphics[width=85mm, height = 30mm]{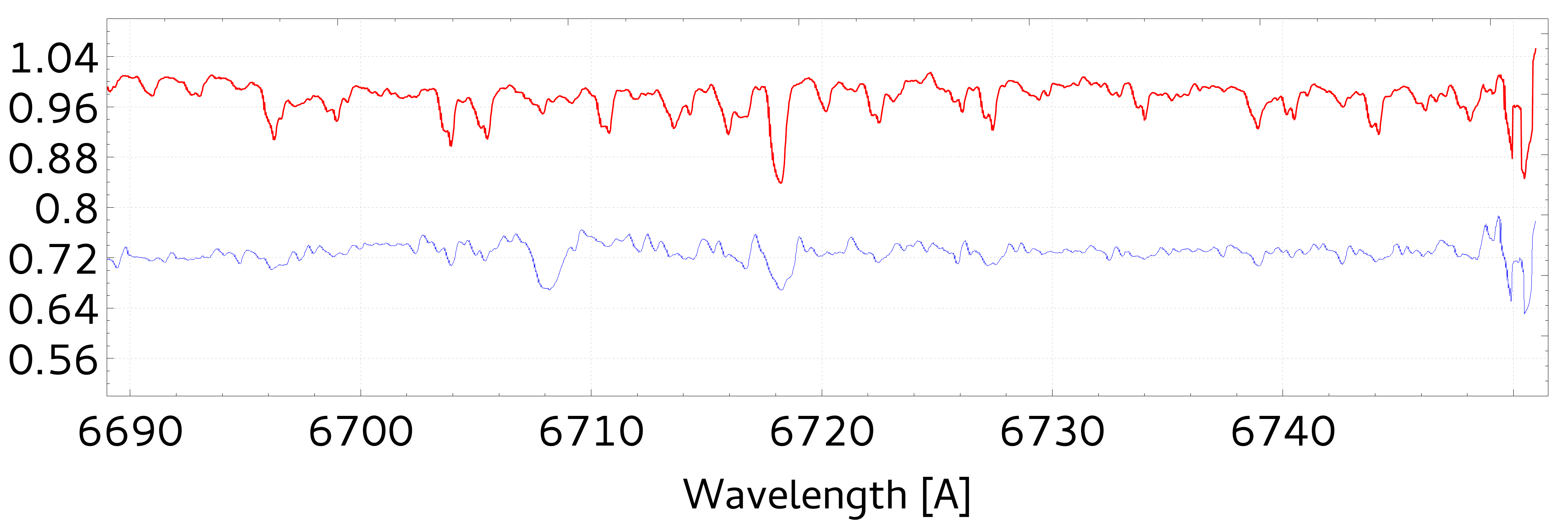}
  \label{Fig17:sfig4}
 \end{subfigure}
\begin{subfigure}{.5\textwidth}
\centering
\caption{Case 5}
 \includegraphics[width=85mm, height = 30mm]{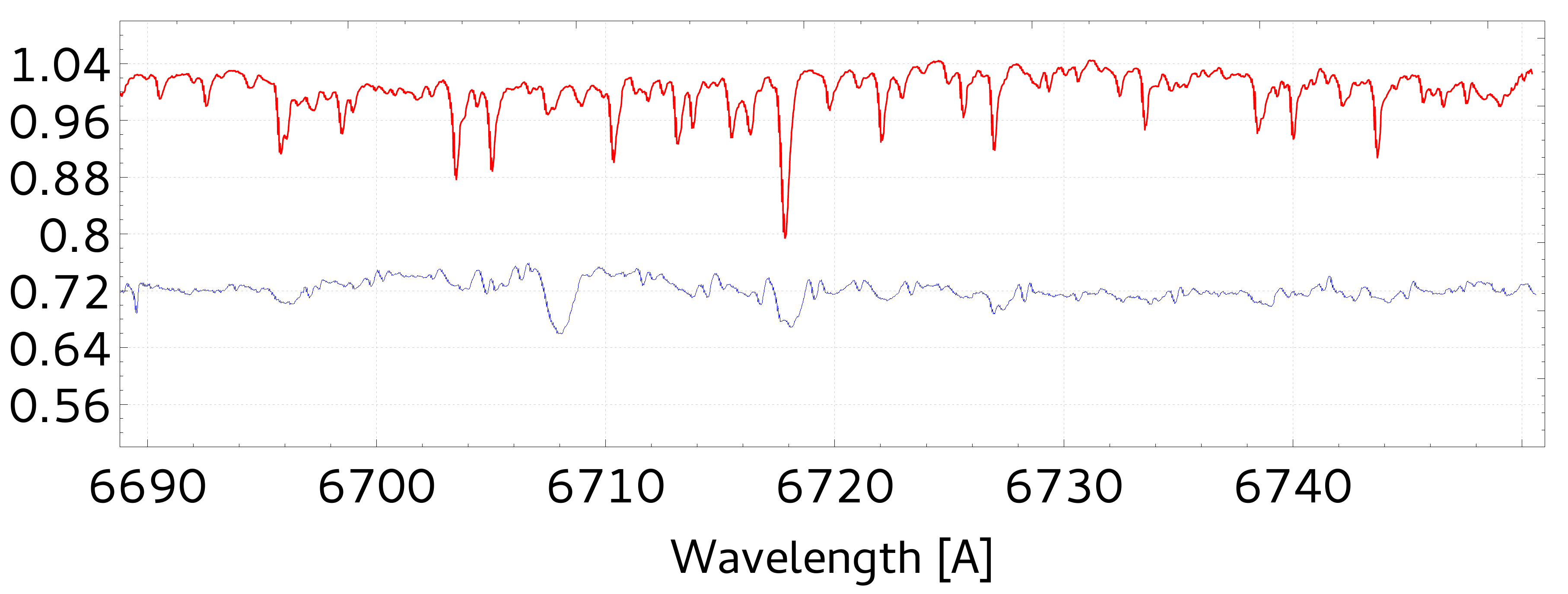}
  \label{Fig17:sfig5}
\end{subfigure}
\begin{subfigure}{.5\textwidth}
  \centering
  \caption{Case 6}
  \includegraphics[width=85mm, height = 30mm]{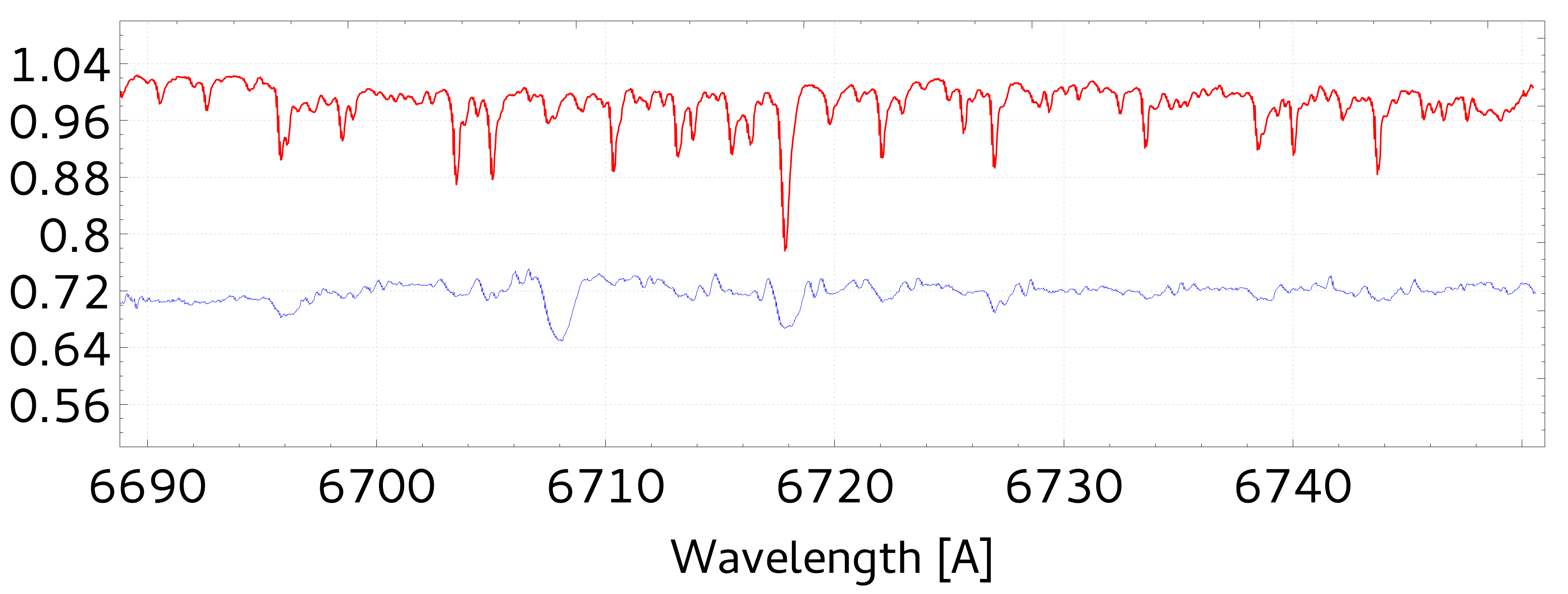}
  \label{Fig17:sfig6}
 \end{subfigure}
\end{center}
 \caption{Results for all six cases of the data from quadrature. See text and Table \ref{tab3} for details.}
  \label{Fig17}
  \end{figure}

The results shown in Fig.~\ref{Fig17} show that data sets which prefer particular phase positions and suffer from many non-independent spectra, degrade the result.
This can be especially seen from cases 2 (spectra from both quadratures) and 4 (all spectra). The other cases show a better separation of line-pairs that are close 
together. However, it is visible that the red wings of these lines show smearing towards the continuum. This comes from regions where line-blending is present and the
SVD is unable to fully distinguish between the two components in these regions. This is even worse if data from both quadratures are used, since all the information
about the line-profile in-between is missing. If precise measurements of line-profiles and equivalent widths are to be made on the resulting spectra, a good phase
coverage is indispensable.

We define a quality factor which relates the number of independent spectra, $N_{IS}$, to the total number of spectra used by their orbital coverage. Let $\widehat{RV}$ 
denote the peak-to-peak value of the radial velocity curve, $S_{RV}$ is the two pixel resolution in the velocity space, $N_S$ the number of all spectra, and $N_{RS}=N_S - N_{IS}$ is the number of reproductions. Thus we define

\begin{equation}
\label{Eq10}
 Q_d \equiv N_{IS} /N_{max} (1-N_{RS}/N_{S}-N_{RS} / N_{max}),
\end{equation}
where $N_{max}=\widehat{RV}/S_{RV}$ is the maximum number of independent spectra.
This factor is at maximum unity as long as redundant measurements are not counted, $N_S \le N_{max}$. The bracket accounts for a statistical overweight of a specific
phase position which leads to a negative $Q_d$. Where independent spectra compensate reproductions, $N_{RS}=N_{IS}$, the factor is zero, i.e., the independent spectra compensate preferred phase positions. 
We list this factor for all six cases in the last column of Table \ref{tab3} and note that for all cases $Q_d < 0$ and case 6 indicates the best result.
Moreover, we note that the quality factor is at 6\,\% for the data set of only five spectra ($N_{IS}=3$) used in Sect. \ref{minimum}.
Hence, this small data set is superior to the larger set that was used here. This can be seen from a comparison of line-separation, e.g., of the
two \ion{Fe}{i} lines blue from the strong \ion{Ca}{i} $\lambda$6717.68{\AA} in the disentangled spectra of the primary (Fig. \ref{Fig17} and Fig. \ref{Fig14:sfig4}, $r=0.83 \cdot 10^{-4}$).
  
  \subsection{The realistic ideal case}
  \label{real}
  
  As shown in the previous two sections, data should be spread homogeneously over at least half the orbital period such that the SVD is able to identify the differential moving contributions. In this case we selected 18 spectra as indicated in Table~\ref{tab2} to get a realistic data set spread over one orbit. The radial  velocities are shown in Fig.~\ref{Fig12:sfig3}. Unlike the data set at quadrature, these data do not show any strong preference in any particular orbit section.
The quality factor $Q_d$ is approximately 20\,\% and indicates much better coverage of the orbit information.
  
\begin{figure}[h!]
 \begin{center}
 \includegraphics[width=85mm]{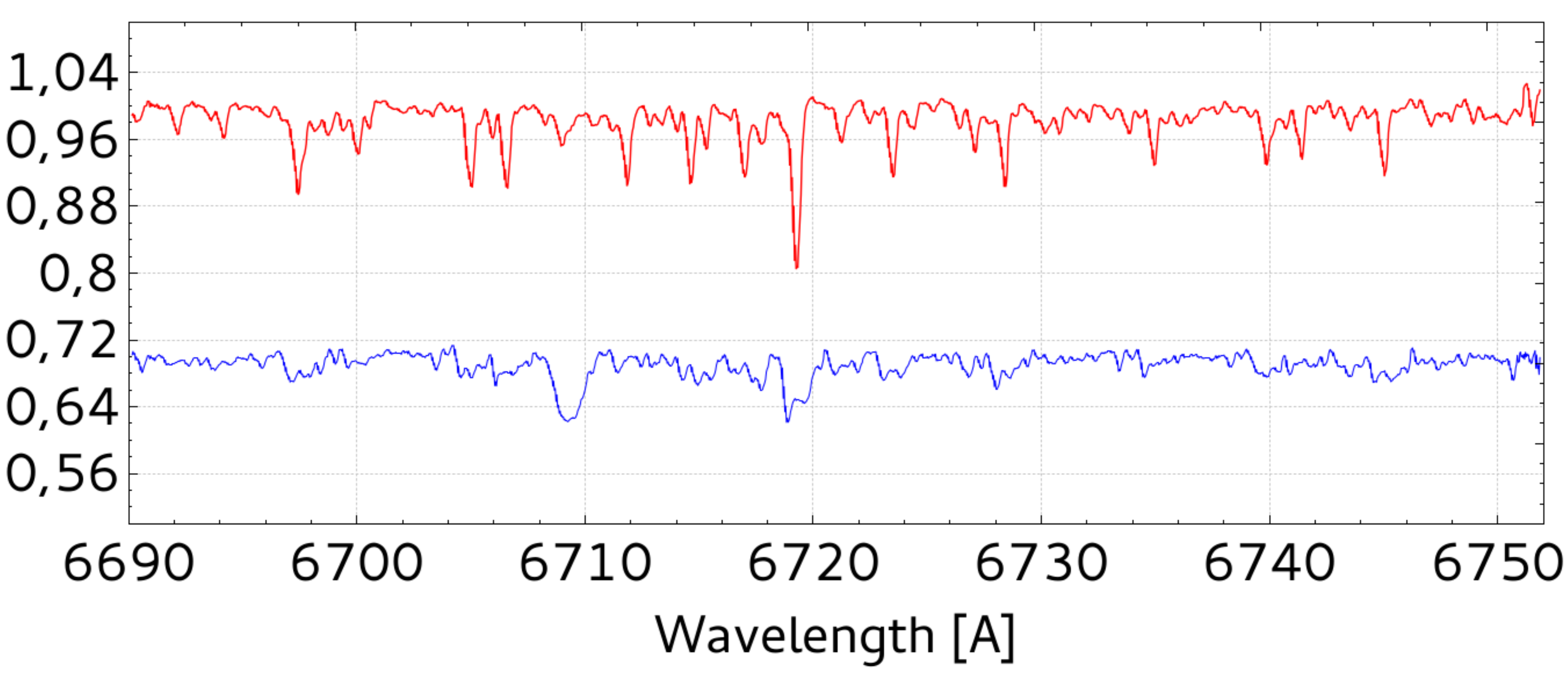}
\end{center}
 \caption{Result of the disentangling of the data indicated in the last column of Table~\ref{tab2}.}
  \label{Fig18}
  \end{figure}
  
The result for this data is shown in Fig.~\ref{Fig18}, where better defined line-wings are visible when compared with the results from the previous section. The narrow lines are now much better separated.

\section{Conclusion}
\label{sec:conclusio}

Using artificial data, we have shown the applicability both of disentangling and of the presented code, in particular to a variety of spectroscopic data from SB1, as well as SB2 systems, including static lines. Additionally, the application to high-resolution data from Capella shows that, with well-sampled small data sets good results can be achieved.
Owing to the broad lines and compared to the primary shallow secondary lines, Capella is a very challenging object.

\begin{figure}[h!]
 \begin{center}
 \includegraphics[width=85mm]{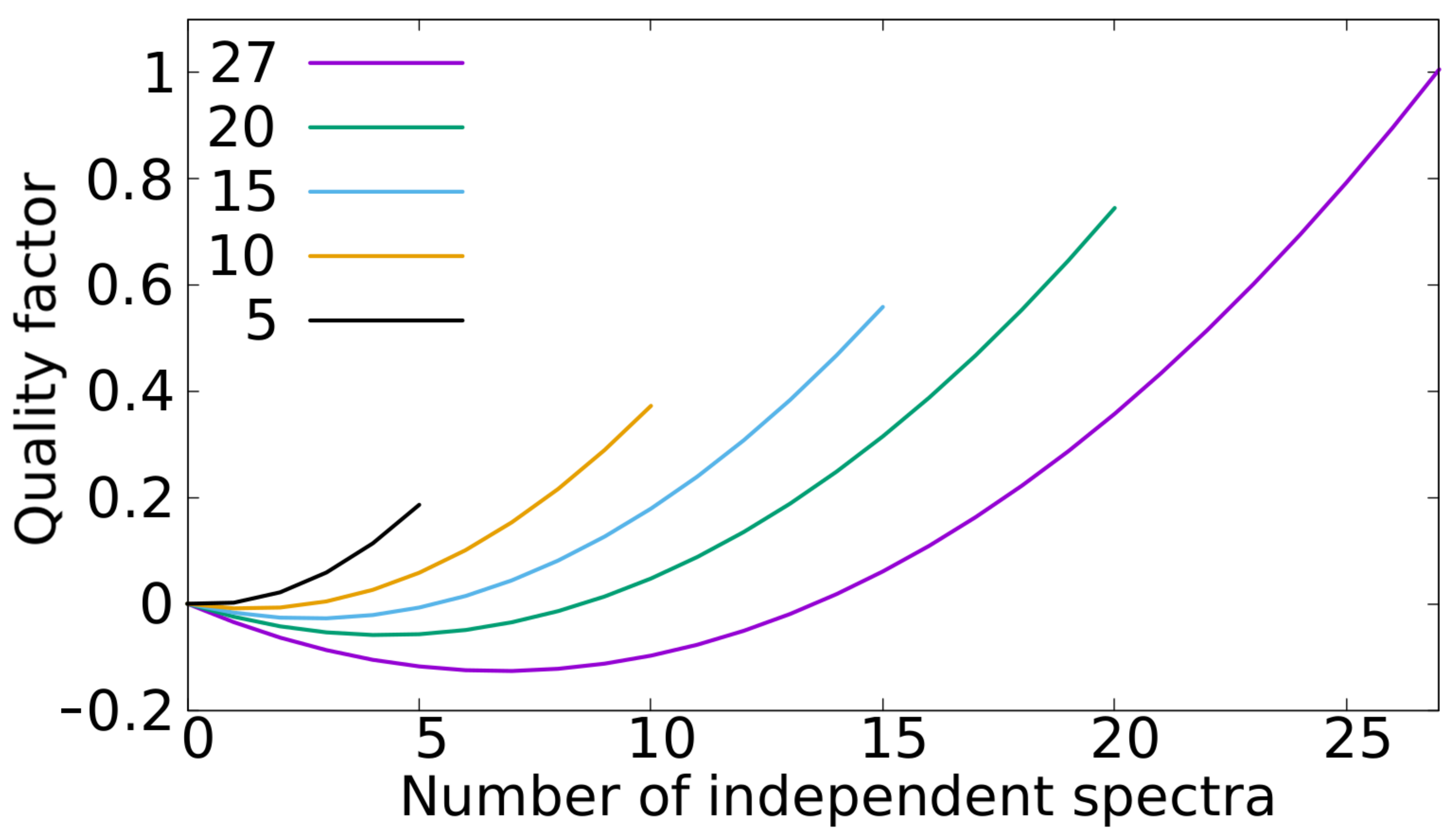}
\end{center}
 \caption{Example of the quality factor in dependence of $N_{IS}$ for $N_{max}=27$ for different $N_S$ as listed in the graph key.}
  \label{Fig19}
  \end{figure}

We conclude that the ideal case for observations is given by a set of spectra spread over one period with a difference of two times the spectrum-sampling in radial velocities. This criterion is based on the Nyquist two-pixel-resolution. It requires only half the data since the criterion given by \citet{1994A&A...281..286S}, where they suggest one spectra per a single velocity bin. However, our tests with Capella have shown that these data are helpful to reducing noise, but they do not significantly contribute to determining the
component spectra (increment of overdetermination). Furthermore, the data should not overweigh a specific phase position. The quality factor given by Eg. \ref{Eq10} should be used to plan observations, especially when the orbit is already known. We plot this relation in Fig. \ref{Fig19} for $N_{max}=27$ and for different $N_S$ in
dependence of $N_{IS}$. It is $N_S=N_{max}$ for the purple curve and if all 27 spectra are independent, i.e., $N_{IS}=N_S=N_{max}$ it yields an $Q_d=1$. For the green curve it is $N_S=20<N_{max}$, i.e., $Q_d<1$ even if all observations are independent, $N_{IS}=N_S=20$.
We also note that all the spectra within the data set need to be of equal quality with respect to S/N and normalisation. A good example is the time-series shown in Fig. \ref{Fig16} where all the spectra are nicely normalised and have identical S/N values.

When there is no initial information on the orbital parameters of a system, we recommended performing a cross-correlation first to get an estimate for the orbital parameters.
The optimisation is then initiated with this orbit, which helps to avoid stagnation of the DSA since it should be already in the vicinity of the best fit. However, optimisation of individual RVs instead of orbital parameters, could be preferable in the case
of the presence of a third component or systematics, e.g., in wavelength calibration. The orbital parameters and the corresponding errors are then found by fitting to the optimised RVs.

Furthermore, \texttt{Spectangular} is able to disentangle telluric lines that are variable in time if relative ratios of the line-depths are given. These ratios
need to be measured for each of the spectra. As long as line-depths can be measured from unblended lines, it is also possible to account for line-depth variability of the component spectra itself. This will also make it possible to use spectra from an eclipse if a light curve of the system is available.
This could help with getting better phase coverage. However, if the uncertainties in the flux ratios are high and the eclipse is short compared to the
orbital period, they can be rejected.

However, despite parallel computing, SVD is a time-consuming procedure. This is underlined when optimisation needs to be performed. Also, the amount of necessary memory is high, since the number of elements to be stored grows quadratically with the spectrum size. Further applications related to time-dependent variability of line-shape rather than depth need to be investigated.

Finally, we note that we are working on the implementation of a third, non-static component. However, we will not couple this to an equation of orbital motion.
Hence, the optimisation will be performed in the case of an SB3 system only on the individual RVs.

\paragraph{Code availability}
We will make the code available under the Apache 2.0 licence on github ASAP. This will include the disentangling program \texttt{Spectangular} as well as \texttt{CroCo} for data preparation.
Furthermore, a support page on the internet and manuals for both programs will be made available.

https://github.com/DPSablowski/croco

https://github.com/DPSablowski/spectangular

\begin{acknowledgements}
We thank the referee for his constructive comments which helped to improve the content and the readability of this paper.
\end{acknowledgements}

\bibliographystyle{aa}
\bibliography{my-library}

\end{document}